\documentclass[superscriptaddress,aps,prb,amsmath,amssymb,twocolumn,letterpaper,
         footinbib,bibnotes,showpacs,reprint,balancelastpage,raggedbottom,longbibliography, %
         citeautoscript,floatfix]{revtex4-2}

\usepackage[dvipsnames,table]{xcolor}
\usepackage[%
  breaklinks,             %
  bookmarks = false, %
  pdfpagemode   = UseNone,%
  pdfstartview  = FitH,     %
  pdfstartpage  = 1,      %
  colorlinks    = true   %
]{hyperref}
\hypersetup{
    linkcolor=RubineRed,          
    citecolor=RoyalBlue,        
    filecolor=Mulberry,      
    urlcolor=RoyalBlue           
}
\usepackage{amssymb}
\usepackage{amsmath}
\usepackage{graphicx}
\usepackage{float}
\usepackage{enumerate}
\usepackage{microtype}
\usepackage{xspace}
\usepackage{siunitx}
\usepackage{upgreek}
\usepackage{bm}
\usepackage{nicefrac}
\usepackage{booktabs}

\usepackage[none]{hyphenat} 
\usepackage{lineno}
\usepackage[utf8]{inputenc}
\DeclareUnicodeCharacter{2212}{-}



\begin{document}

\author{Yongjin Shin}
\email{yongjin.shin@dankook.ac.kr}
\affiliation{Department of Semiconductor Convergence Engineering, Dankook University, Yongin-si 16890, Gyeonggi-do, South Korea}

\def\degree{$^\circ$\xspace}
\newcommand\Tstrut{\rule{0pt}{2.6ex}}         
\newcommand\Bstrut{\rule[-0.9ex]{0pt}{0pt}}   
\newcommand\LTO{LaTaO$_{4}$\xspace}
\newcommand\LNO{LaNbO$_{4}$\xspace}
\newcommand\RTO{$R$TaO$_{4}$\xspace}
\newcommand\RNO{$R$NbO$_{4}$\xspace}
\newcommand\Ccm{$\mu\mathrm{C/cm}^2$\xspace}
\newcommand\Eform{$\Delta E_\mathrm{form}$\xspace}
\newcommand\PathOrtho{$Cmc2_1 \rightarrow P2_1$\xspace}
\newcommand\PathMono{$P2_1/c \rightarrow P2_1$\xspace}

\title{\boldmath Metastable polar order and phase competition in Carpy-Galy LaTaO$_4$}

\begin{abstract}
Carpy-Galy LaTaO$_4$ undergoes a structural sequence linking the antipolar monoclinic $P2_1/c$ ground state and the polar orthorhombic $Cmc2_1$ phase through an incommensurately modulated regime.
First-principles calculations identify a metastable polar monoclinic $P2_1$ phase as a commensurate intermediate in the phase-transition landscape.
As a common subgroup of $P2_1/c$ and $Cmc2_1$, $P2_1$ provides a symmetry-connected route between the established polymorphs.
Mode-resolved energy surfaces for these symmetry-breaking pathways show that coupling between the respective primary order parameters and an isosymmetric relaxation stabilizes the $P2_1$ minimum along both pathways. 
Consequently, the $P2_1/c\rightarrow P2_1$ transformation has a finite energy barrier, while a restricted soft-mode model retains competing $P2_1$ and $Cmc2_1$ basins over an illustrative range of harmonic stiffness.
Lattice-metric and neutron-diffraction comparisons indicate a shared displacement character between the calculated commensurate state and the experimental IC-o modulation, supporting the use of $P2_1$ as a commensurate structural reference for IC-o.
Finally, composition-dependent energetics suggest chemical routes for tuning polar-antipolar competition.
\end{abstract}

\maketitle

\section{Introduction}

\begin{figure}[b]
\centering
\includegraphics[width=0.99\columnwidth]{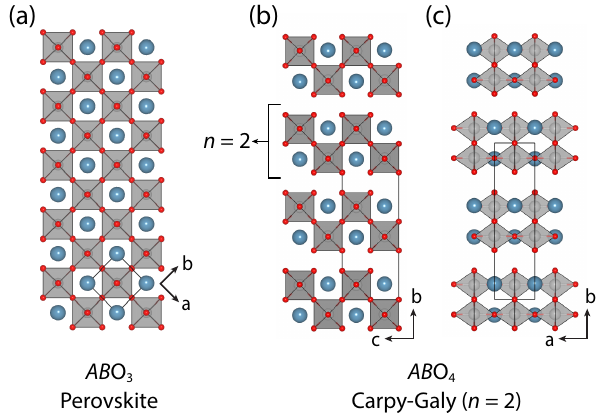}
\caption{(a) Ideal $AB$O$_3$ perovskite. (b,c) The $n=2$ Carpy-Galy topology viewed along $a$ and $c$, respectively. Adjacent perovskite slabs are translated along $a$ by half an octahedral unit upon stacking. Blue, grey, and red spheres represent $A$, $B$, and O atoms.}
\label{fig:Schematic_Structure}
\end{figure}

Geometric ferroelectricity provides a route to polar order in which lattice connectivity and collective distortions generate polarization without relying on a particular electronic instability \cite{HIF:Benedek2011,Shin/Galli2023,JHJang/SYChoi2025,Lu/Rondinelli2016,Gradauskaite2026,Spaldin2019}.
The reliance on atomic structure instead of electronic configurations broadens the chemical space available for functional polar materials \cite{Iniguez2011,NunezValdez/Spaldin2019_Origin}. Layered perovskite derivatives, such as Ruddlesden--Popper oxides \cite{HIF:Benedek2011,Lu/Rondinelli2016,Shin/Rondinelli2017}, exemplify this mechanism because truncation of the octahedral network introduces structural degrees of freedom absent from the three-dimensional perovskite.
Carpy-Galy structures form a less-explored family of layered perovskite derivatives distinguished by slabs truncated parallel to the $(110)$ planes of the parent perovskite \cite{Galy1974,Carpy1972,Shin/Rondinelli2017}.
By contrast, the octahedral networks of more familiar layered oxides are truncated parallel to the $(001)$ planes. Oxygen termination of the finite slabs in Carpy-Galy structure gives the general formula $A_nB_n$O$_{3n+2}$, where $n$ denotes the number of perovskite layers within each slab (\autoref{fig:Schematic_Structure}).
For even $n$, the slab topology can prevent complete cancellation of rotation-induced displacements at the terminating layers, allowing cooperative octahedral rotations to generate a net polarization \cite{Liu/Chen2016_Topological,NunezValdez/Spaldin2019_Origin,Iniguez2011,Ederer/Spaldin2006,Valdez/Spaldin2016,Shin/Galli2023}.

\begin{figure*}[t]
\centering
\includegraphics[width=1.8\columnwidth]{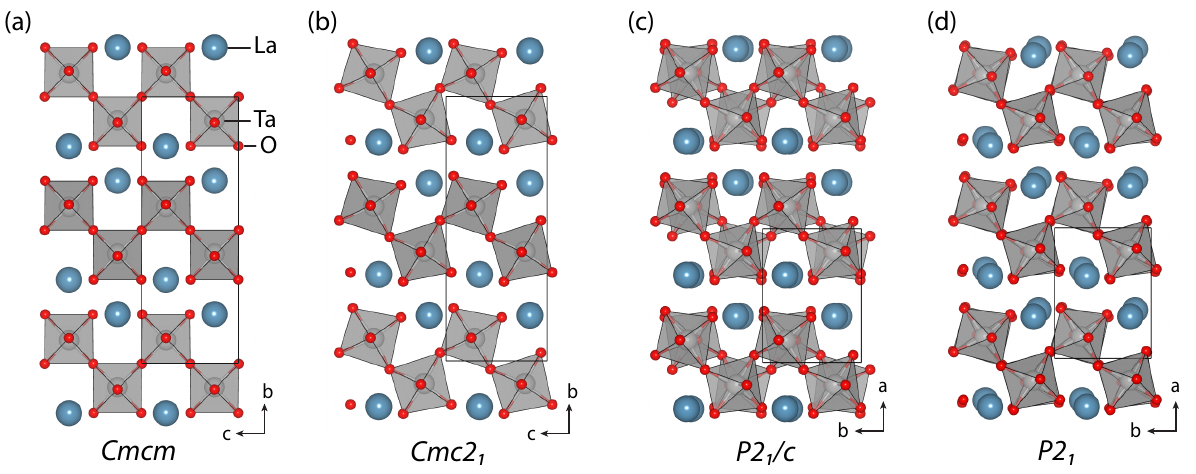}
\caption{Calculated polymorphs of Carpy-Galy LaTaO$_4$: (a) nonpolar orthorhombic $Cmcm$, (b) polar orthorhombic $Cmc2_1$, (c) antipolar monoclinic $P2_1/c$, and (d) polar monoclinic $P2_1$.
Relative to the unrotated $Cmcm$ parent, $Cmc2_1$ contains the in-phase octahedral-rotation pattern, $P2_1/c$ contains the compensating out-of-phase pattern, and $P2_1$ contains nearly in-phase rotations with alternating amplitudes. Blue and red spheres represent La and O atoms, respectively.}
\label{fig:Structures_front}
\end{figure*}

LaTaO$_4$ is a representative $n=2$ Carpy-Galy compound and undergoes a temperature-dependent structural sequence from the antipolar monoclinic $P2_1/c$ ground state to the polar orthorhombic $Cmc2_1$ phase above approximately 520 K [\autoref{fig:Structures_front}(b,c)] \cite{Abreu2017,Cordrey/Lightfoot2015,Howieson/Morrison2020_Incommensurate,Howieson/Carpenter2021}.
Interestingly, the temperature-dependent sequence contains first-order phase boundaries and structurally distinct intermediate regimes.
For example, neutron diffraction identifies an incommensurately modulated orthorhombic (IC-o) regime with a modulation vector $q\simeq(0.46,0,0)$ \cite{Howieson/Morrison2020_Incommensurate,Howieson/Carpenter2021}.
More recently, in situ electron-diffraction measurements collected during heating resolved a monoclinic phase, denoted $m'$, over a narrow temperature interval between $P2_1/c$ and IC-o, and the accompanying analysis assigned $P2_1$ symmetry to the phase \cite{Wu/Howieson/Ma2026_In_situ}.
Despite the experimentally resolved complexity, previous first-principles studies focused mainly on the established $Cmcm$, $Cmc2_1$, and $P2_1/c$ polymorphs \cite{Liu/Chen2016_Topological,Cui/Tian2024}.
The energetic accessibility of an intermediate structure and the microscopic relationship between a commensurate intermediate and IC-o therefore remain unresolved.

In this work, first-principles calculations identify a metastable polar monoclinic $P2_1$ minimum whose 0 K energy lies between the energies of the $P2_1/c$ ground state and the high-temperature $Cmc2_1$ phase. 
The common-subgroup relationship provides a symmetry-connected route between the established polymorphs at lower energy than the route through $Cmcm$. Mode-resolved energy surfaces characterize the $P2_1/c\rightarrow P2_1$ segment as a proper-ferroelectric transformation with a finite barrier, whereas the $Cmc2_1\rightarrow P2_1$ pathway begins with the zone-boundary $S_2$ instability.
Along both pathways, coupling between the respective order parameters and an isosymmetric relaxation plays a key role in stabilizing the $P2_1$ minimum. 
A restricted soft-mode model retains distinct $P2_1$ and $Cmc2_1$ basins over a finite range of harmonic stiffness chosen for illustration, consistent with metastability and possible phase competition. In addition, lattice-metric and neutron-diffraction comparisons indicate a shared $S_2$-dominated displacement character between the calculated $P2_1$ structure and IC-o, supporting the use of $P2_1$ as a $q=1/2$ zone-boundary commensurate reference. 
Finally, a survey of $R$TaO$_4$ and $R$NbO$_4$ ($R=$ La, Ce, Pr, and Nd) identifies compositional trends in polymorph competition and suggests chemical routes for tuning polar-antipolar phase behavior.

\section{Computational Methods}

Density functional theory (DFT) calculations were performed with the Vienna Ab initio Simulation Package (\textsc{vasp}) \cite{Kresse1996VASP,Kresse1999}, using the revised Perdew--Burke--Ernzerhof functional for solids (PBEsol) \cite{PBE,PbESol} and projector-augmented-wave (PAW) potentials \cite{Blochl1994}. The valence-electron configurations were La ($5s^2 5p^6 5d^1 6s^2$), Ce ($5s^2 5p^6 5d^1 6s^2$), Pr ($5s^2 5p^6 5d^1 6s^2$), Nd ($5s^2 5p^6 5d^1 6s^2$), Nb ($4s^2 4p^6 4d^4 5s^1$), Ta ($5s^2 5p^6 5d^4 6s^1$), and O ($2s^2 2p^4$). The calculations used a plane-wave energy cutoff of 550\,eV, the tetrahedron method \cite{Bloch1994Tetrahedron}, and $\Gamma$-centered $k$-point meshes. To maintain comparable sampling densities, meshes of $8\times2\times6$ and $5\times5\times5$ were used for the orthorhombic ($Cmcm$ and $Cmc2_1$) and monoclinic ($P2_1/c$ and $P2_1$) cells, respectively. The cell parameters and atomic coordinates were relaxed until all residual forces were below 3~meV\,\AA$^{-1}$.

Spontaneous polarizations were evaluated using the Berry-phase formulation of the modern theory of polarization \cite{Spaldin2012,Resta/Vanderbilt2007}. Phonon dispersion relations were calculated using the finite-displacement method implemented in \textsc{phonopy} with $2\times2\times2$ supercells \cite{Phonopy}. The resulting phonon dispersions are provided in the Supplemental Material \cite{Supp} (see also Refs.~\onlinecite{Shin/Lee2016,Kim/Shin2026,Huang/Rondinelli2016} therein).

Symmetry-adapted mode decompositions were performed with \textsc{isodistort} \cite{Campbell:ISODISTORT,ISOTROPY,FINDSYM}, and minimum-energy trajectories on the two-dimensional Landau surfaces were determined by string relaxation. The barrier along \PathMono was also evaluated using six-image solid-state nudged elastic-band (SS-NEB) calculations \cite{Henkelman2012_SSNEB}. The SS-NEB calculations used cell parameters interpolated between those of the $P2_1/c$ and $P2_1$ endpoints and applied the same force-convergence threshold as the conventional structural relaxations.

Neutron-diffraction patterns were simulated with \textsc{pymatgen} \cite{Ong2013Pymatgen,DeGraefMcHenry2007}, and the experimental patterns used for comparison were taken from Ref.~\onlinecite{Howieson/Morrison2020_Incommensurate}.

\section{Results and Discussion}

\begin{figure}[h]
\centering
\includegraphics[width=0.90\columnwidth]{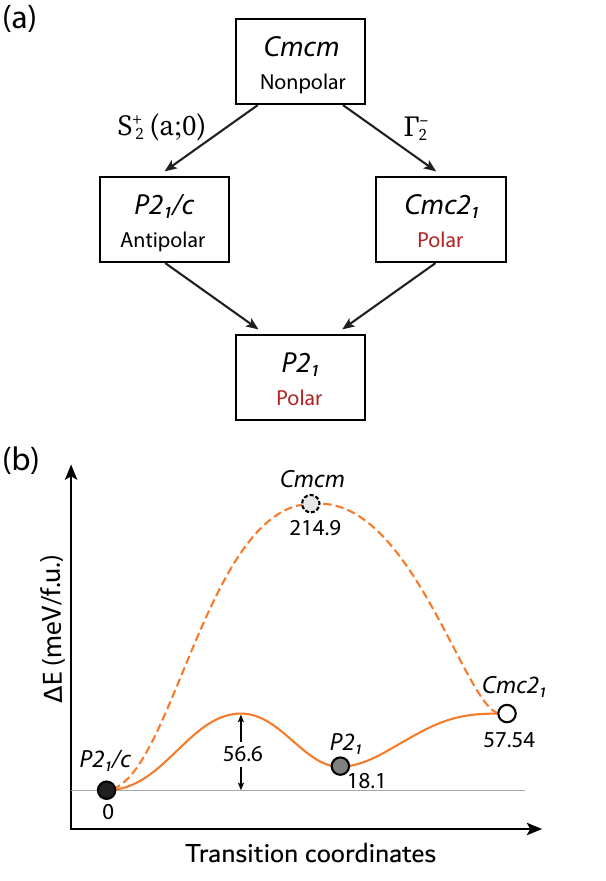}
\caption{(a) Group-subgroup relations among the $Cmcm$, $Cmc2_1$, $P2_1/c$, and $P2_1$ phases. (b) Calculated 0 K energy profiles for the transformation from $P2_1/c$ to $Cmc2_1$ through either $P2_1$ or $Cmcm$. Energies are given relative to $P2_1/c$, and the vertical arrow indicates the SS-NEB \cite{Henkelman2012_SSNEB} barrier from $P2_1/c$ to $P2_1$.}
\label{fig:Group_tree}
\end{figure}

\subsection{Competing Carpy-Galy polymorphs}

The three established Carpy-Galy polymorphs of \LTO, $Cmcm$, $Cmc2_1$, and $P2_1/c$, are distinguished primarily by the rotation patterns of the TaO$_6$ octahedra [\autoref{fig:Structures_front}(a--c)] \cite{Liu/Chen2016_Topological,Cui/Tian2024}. 
The highest-symmetry $Cmcm$ phase contains no octahedral rotations, although La cations shift outward from the perovskite slabs relative to the ideal truncated cubic-perovskite structure [\autoref{fig:Schematic_Structure}(b,c) and \autoref{fig:Structures_front}(a)]. Condensation of in-phase rotations about the $a$ axis produces polar $Cmc2_1$. 
The rotation distortions orient the open sides of the La pockets along $-c$, generating coherent La displacements and a net polarization [\autoref{fig:Structures_front}(b)]. 
By comparison, monoclinic $P2_1/c$ contains pronounced out-of-phase rotations about the $a$ axis. Alternation of the resulting La-pocket distortions along $a$ cancels the local dipoles and produces an antipolar structure [\autoref{fig:Structures_front}(c)].

At 0 K, $P2_1/c$ is the ground state, whereas $Cmc2_1$ and $Cmcm$ lie 57.54 and 214.9 meV/f.u. higher, respectively (\autoref{table:LTO_Lattice_Energy}).
The PBEsol energy ordering agrees with previous calculations \cite{Liu/Chen2016_Topological}, and the corresponding mode-resolved energy profiles are provided in the Supplemental Material \cite{Supp}. 
Both $P2_1/c$ and $Cmc2_1$ are subgroups of $Cmcm$, but neither structure is a subgroup of the other [\autoref{fig:Group_tree}(a)]. 
Therefore, a direct continuous transition between polar $Cmc2_1$ and antipolar $P2_1/c$ is not symmetry-allowed, consistent with the first-order character of the experimental phase sequence \cite{Howieson/Morrison2020_Incommensurate}. 
Although a pathway through unrotated $Cmcm$ remains allowed by symmetry, $Cmcm$ lies 214.9 meV/f.u. above $P2_1/c$ and 157.4 meV/f.u. above $Cmc2_1$, making the high-symmetry route energetically unfavorable [\autoref{fig:Group_tree}(b)].

By contrast, the polar monoclinic $P2_1$ phase shown in \autoref{fig:Structures_front}(d) is a common subgroup of $P2_1/c$ and $Cmc2_1$ and provides a lower-energy symmetry-connected pathway [\autoref{fig:Group_tree}(a)].
Condensation of the zone-boundary $S_2$ instability at $q=(1/2,0,0)$ from $Cmc2_1$, followed by secondary zone-center relaxations, produces the structure shown in \autoref{fig:Structures_front}(d).
Although both in- and out-of-phase rotations are symmetry-allowed in the $P2_1$ structure of \LTO, the optimized geometry contains a nearly in-phase pattern of neighboring octahedral rotations, with amplitudes that alternate along the $c$ direction. 
The associated La displacements alternate along $c$, partially canceling the $b$-axis dipole components. 
Consequently, the calculated spontaneous-polarization magnitude decreases from 36.1~\Ccm\ in $Cmc2_1$ to 16.2~\Ccm\ in $P2_1$ \cite{Supp}. The $P2_1$ minimum lies 18.1 meV/f.u. above the $P2_1/c$ ground state and 39.4 meV/f.u. below $Cmc2_1$. The corresponding phonon analysis supports the dynamical stability of the metastable minimum \cite{Supp}.
The same energetic ordering is obtained with r$^2$SCAN and HSE06, supporting the robustness of the intermediate energetic position of $P2_1$ against the choice of exchange–correlation functional \cite{Supp}.

The group-subgroup relationship and associated distortion patterns resemble those found in $n=4$ Carpy-Galy structures \cite{Iniguez2011,NunezValdez/Spaldin2019_Origin,Bruyer/Sayede2010,Gradauskaite2025}.
Although this specific $P2_1$ phase has not been identified in $n=2$ Carpy-Galy compounds, previous studies have considered $P2_1$ symmetry in \LTO.
For example, Wu et al.\ observed a transient phase denoted $m'$ and assigned $P2_1$ symmetry to the phase \cite{Wu/Howieson/Ma2026_In_situ}. However, unlike the $P2_1$ phase in \autoref{fig:Structures_front}(d), the $c$ lattice parameter of the $m'$ phase is approximately half that of the present $P2_1$ structure, indicating that $m'$ and the calculated $P2_1$ phase are structurally distinct.

Liu et al.\ theoretically examined a $P2_1$ structure generated from the $S$-point instability of the $Cmc2_1$ phase through a symmetry pathway equivalent to that considered here \cite{Liu/Chen2016_Topological}. The \textit{ab initio} calculations in that study found that the imposed $P2_1$ structure relaxed to $P2_1/c$ and concluded that $P2_1$ was not a stable variant.
The different structural relaxation outcomes are attributed to the initial coordinates and the energy barrier depicted in \autoref{fig:Group_tree}(b).
Due to the energy barrier, a starting $P2_1$ configuration close to $P2_1/c$ relaxes into the ground-state basin, whereas condensation of $S_2$ from a configuration close to $Cmc2_1$ can reach the metastable local minimum.
The following subsections quantify the pathways and energetic barriers separating these basins.

The calculated $P2_1$ minimum consequently plays two complementary roles in the phase-transition landscape of \LTO.
First, $P2_1$ provides a metastable intermediate phase along a lower-energy pathway between $P2_1/c$ and $Cmc2_1$.
Second, the $P2_1$ phase motivates comparison with the experimentally observed IC-o phase, which also occurs as an intermediate during the polar-antipolar phase-transition sequence.
Although their translational periodicities differ, the local structural similarities discussed below support using the calculated $P2_1$ state as a commensurate reference for IC-o.

\begin{table}[h]
\caption{\label{table:LTO_Lattice_Energy}
Calculated 0 K properties of the LaTaO$_4$ polymorphs. $\Delta E$ is the total energy relative to the $P2_1/c$ ground state, $E_g$ is the PBEsol Kohn-Sham band gap, and $P$ is the magnitude of the spontaneous polarization obtained from the Berry-phase formalism.
}
\centering
\begin{ruledtabular}
\begin{tabular}{r r r r}
     Space group    & $\Delta E$ (meV/f.u.)  & $E_g$ (eV) & $P$ ($\mu\text{C/cm}^2$)\Bstrut \\ 
     \hline \Tstrut    
$Cmcm$ (No. 63)      & 214.9  &  2.89 & 0  \\     
$Cmc2_1$ (No. 36)    & 57.54  &  3.32 & 36.1  \\
$P2_1/c$ (No. 14)    & 0      &  3.44 & 0  \\
$P2_1$ (No. 4)       & 18.1   &  3.33 & 16.2  \\
\end{tabular}
\end{ruledtabular}
\end{table}

\begin{figure*}[t]
\centering
\includegraphics[width=0.99\textwidth]{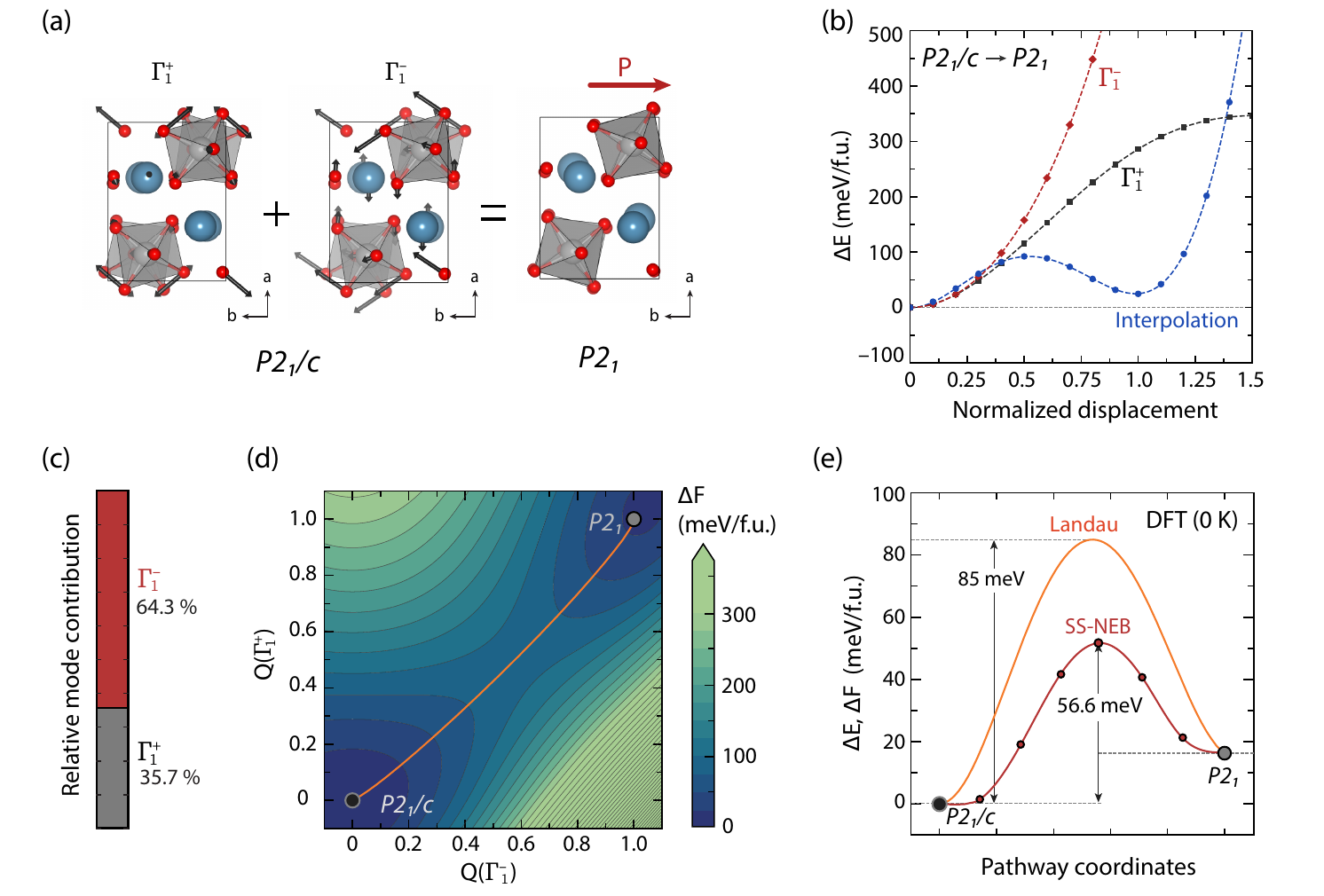}\vspace{-0.4em}
\caption{(a) Symmetry-adapted $\Gamma_1^+$ and $\Gamma_1^-$ distortions connecting $P2_1/c$ and $P2_1$. (b) Calculated energies of the individual modes and their combined linear interpolation, referenced to $P2_1/c$. (c) Relative mode contributions to the total distortion. (d) Two-mode Landau energy landscape and string-relaxed minimum-energy path. (e) Energy profiles along the Landau ($\Delta F$) and six-image SS-NEB ($\Delta E$) paths, with barriers of 84.6 and 56.6 meV/f.u., respectively.}
\label{fig:Str_P21c-P21}
\end{figure*}

\subsection{\boldmath Transition from $P2_1/c$ to the metastable $P2_1$ state}
\label{subsec:P21c_P21}

With $P2_1$ established as a metastable local minimum, the lower-energy symmetry-connected route is first examined along \PathMono. 
Relative to the $P2_1/c$ parent, the structural difference decomposes into the zone-center $\Gamma_1^-$ and $\Gamma_1^+$ irreducible representations [\autoref{fig:Str_P21c-P21}(a)]. 
The polar $\Gamma_1^-$ mode removes inversion symmetry and generates polarization along the unique monoclinic axis ($b$), whereas $\Gamma_1^+$ preserves the parent symmetry. 
Thus, $\Gamma_1^-$ is the sole symmetry-breaking order parameter, classifying the $P2_1/c\rightarrow P2_1$ transformation as a proper ferroelectric transition despite the accompanying isosymmetric relaxation.

Structurally, the two modes modify the octahedral rotations in complementary ways. 
The $\Gamma_1^+$ distortion changes the amplitudes of the out-of-phase rotations already present in $P2_1/c$, whereas $\Gamma_1^-$ drives neighboring octahedra toward in-phase rotation. 
Combined condensation produces the nearly in-phase rotation pattern with unequal amplitudes found in $P2_1$. 
Although $\Gamma_1^-$ makes the larger contribution, $\Gamma_1^+$ accounts for 35.7\% of the total distortion, demonstrating the substantial structural role of the isosymmetric relaxation [\autoref{fig:Str_P21c-P21}(c)].

As shown in \autoref{fig:Str_P21c-P21}(b), condensing either mode separately raises the energy relative to $P2_1/c$, whereas the combined linear interpolation reaches a second local minimum at $P2_1$.
To quantify the coupled energy surface, the calculated energies were fitted to the following Landau formulation:
\begin{equation}
\begin{split}
F(\eta, Q) = F_0 & + A_2 \eta^2 + A_4 \eta^4  \\
     & + B_2 Q^2 + B_3 Q^3 + B_4 Q^4 \\
&+ C_{21} \eta^2 Q + C_{22} \eta^2 Q^2.
\end{split}
\label{eq:Landau}
\end{equation}
Here, $(\eta,Q)=(Q_{\Gamma_1^-},Q_{\Gamma_1^+})$ for the $P2_1/c\rightarrow P2_1$ pathway and $(\eta,Q)=(Q_{S_2},Q_{\Gamma_1})$ for the $Cmc2_1\rightarrow P2_1$ pathway. The coordinate $Q$ describes the totally symmetric relaxation of the corresponding parent structure. The fitted coefficients are listed in \autoref{table:Landau}. All amplitudes are normalized to the relaxed $P2_1$ structure.
In \autoref{eq:Landau}, odd powers of $Q$ are symmetry allowed because $Q$ is totally symmetric.
Notably, the negative cubic coupling term $C_{21}Q_{\Gamma_1^-}^2Q_{\Gamma_1^+}$, with $C_{21}=-1.7444$ eV/f.u., lowers the energy as the polar and isosymmetric distortions develop together.
This coupling term produces the two-minimum landscape connecting $P2_1/c$ and $P2_1$ in \autoref{fig:Str_P21c-P21}(d).

The coupled landscape next provides a basis for comparing the restricted two-mode trajectory with a more complete transformation path. 
String relaxation on the fitted surface gives a barrier of 84.6 meV/f.u. relative to $P2_1/c$.
By comparison, the six-image SS-NEB calculation reduces the barrier to 56.6 meV/f.u. because the method permits atomic relaxations beyond the two-mode subspace [\autoref{fig:Str_P21c-P21}(e)].
The local $P2_1$ minimum and finite energy barrier generated by higher-order coupling are consistent with the hysteresis of the first-order low-temperature transformation \cite{Howieson/Carpenter2021,Howieson/Morrison2020_Incommensurate}.

The barrier height of 56.6 meV/f.u. between $P2_1/c$ and $P2_1$ is close to the static energy difference of 57.54 meV/f.u. between $Cmc2_1$ and $P2_1/c$.
Although the calculated energy landscape does not explicitly include finite-temperature effects,
the comparable energy values place the energetic cost of leaving the $P2_1/c$ basin on the same scale as the energy offset of $Cmc2_1$. 
The comparison reinforces the relevance of the metastable $P2_1$ state and possible phase competition within the polar-antipolar transition landscape.

\begin{figure*}[t]
\centering
 \includegraphics[width=0.99\textwidth]{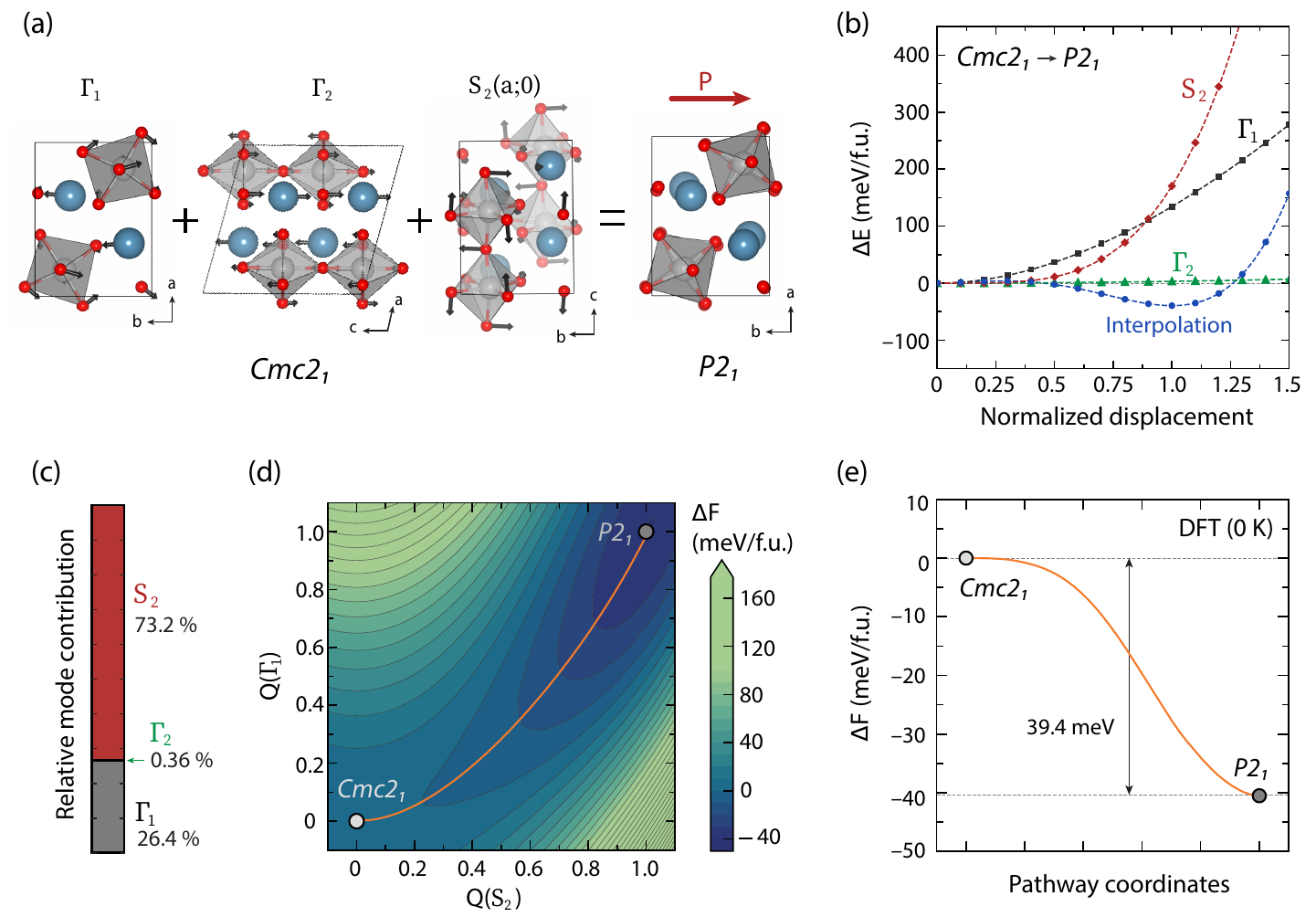}\vspace{-0.4em}
\caption{(a) Symmetry-adapted $\Gamma_1$, $\Gamma_2$, and $S_2$ distortions connecting $Cmc2_1$ and $P2_1$. (b) Calculated energies of the individual modes and their combined linear interpolation, referenced to $Cmc2_1$. (c) Relative mode contributions to the total distortion. (d) Two-mode Landau energy landscape and string-relaxed minimum-energy path. (e) Energy profile along the minimum-energy path, with $P2_1$ lying 39.4 meV/f.u. below $Cmc2_1$.}
\label{fig:Str_Cmc21-P21}
\vspace{0.5em}
\end{figure*}

\begin{table}[h]
\caption{\label{table:Landau}
Fitted Landau coefficients for the $P2_1/c\rightarrow P2_1$ and $Cmc2_1\rightarrow P2_1$ pathways. For the two pathways, the normalized primary order parameter $\eta$ corresponds to $Q_{\Gamma_1^-}$ and $Q_{S_2}$, respectively. The corresponding root-mean-square errors in $F$ are 1.1 and 0.8 meV/f.u.
}
\centering
\setlength{\tabcolsep}{10pt}
\begin{tabular}{c| c c }
\hline \hline
        & \PathMono  & \PathOrtho \Bstrut \\ 
     \hline \Tstrut 
$\eta$        & $Q_{\Gamma_1^-}$       & $Q_{S_2}$    \\     
$F_0$         & 0.0051                & 0.0004          \\
$A_2$         & 0.5347                & -0.0013         \\
$A_4$         & 0.2316                & 0.1682          \\
$B_2$         & 0.6857                & 0.1572          \\
$B_3$         & -0.5490               & -0.0245         \\
$B_4$         & 0.1375                & 0.0001          \\
$C_{21}$      & -1.7444               & -0.4410         \\
$C_{22}$      & 0.7200                & 0.1009          \\
\hline \hline
\end{tabular}  
\end{table}

\subsection{\boldmath Transition from $Cmc2_1$ to the commensurate $P2_1$ reference}

The $Cmc2_1\rightarrow P2_1$ pathway decomposes into the $\Gamma_1$, $\Gamma_2$, and zone-boundary $S_2$ distortions of the $Cmc2_1$ parent [\autoref{fig:Str_Cmc21-P21}(a)].
Because $S_2$ carries the zone-boundary wave vector required by the $P2_1$ periodicity, $Q_{S_2}$ is the primary order parameter and accounts for 73.2\% of the total distortion [\autoref{fig:Str_Cmc21-P21}(c)].
The isosymmetric $\Gamma_1$ mode contributes a further 26.4\%, whereas $\Gamma_2$ contributes only 0.36\% and changes the energy by less than 3 meV/f.u. over the sampled amplitude range [\autoref{fig:Str_Cmc21-P21}(b,c)].
Given the negligible role of the $\Gamma_2$ mode, the energy analysis retains $S_2$ and $\Gamma_1$ as the active coordinates.

With $\eta=Q_{S_2}$ and $Q=Q_{\Gamma_1}$, the $Cmc2_1\rightarrow P2_1$ energy surface is fitted to \autoref{eq:Landau}, and the resulting coefficients are listed in \autoref{table:Landau}.
Neither isolated mode reproduces the energy lowering of the coupled pathway [\autoref{fig:Str_Cmc21-P21}(b)].
For $S_2$, the small negative quadratic coefficient $A_2=-0.0013$ eV/f.u. describes only a shallow, nearly negligible instability.
The larger stabilization of $P2_1$ along \PathOrtho, as along \PathMono, arises from the negative cubic coupling term $C_{21}Q_{S_2}^2Q_{\Gamma_1}$, with $C_{21}=-0.4410$ eV/f.u., which produces the diagonal valley in \autoref{fig:Str_Cmc21-P21}(d).

The minimum-energy trajectory on the fitted surface departs from the direct linear interpolation, showing that the two distortions develop at different stages [\autoref{fig:Str_Cmc21-P21}(d)]. 
Near $Cmc2_1$, $S_2$ develops first because $A_2<0$, whereas the positive $B_2$ makes $\Gamma_1$ stable in isolation. 
As $S_2$ grows, the cubic coupling term ($C_{21}Q_{S_2}^2Q_{\Gamma_1}$) induces the isosymmetric $\Gamma_1$ relaxation. 
The coupled evolution of the two modes stabilizes the $P2_1$ minimum at 39.4 meV/f.u. below $Cmc2_1$ [\autoref{fig:Str_Cmc21-P21}(e)].
Thus, $P2_1$ forms a metastable intermediate on the calculated 0 K landscape, with an energy between those of $P2_1/c$ and $Cmc2_1$.

\begin{figure*}[t]
\centering
 \includegraphics[width=0.99\textwidth]{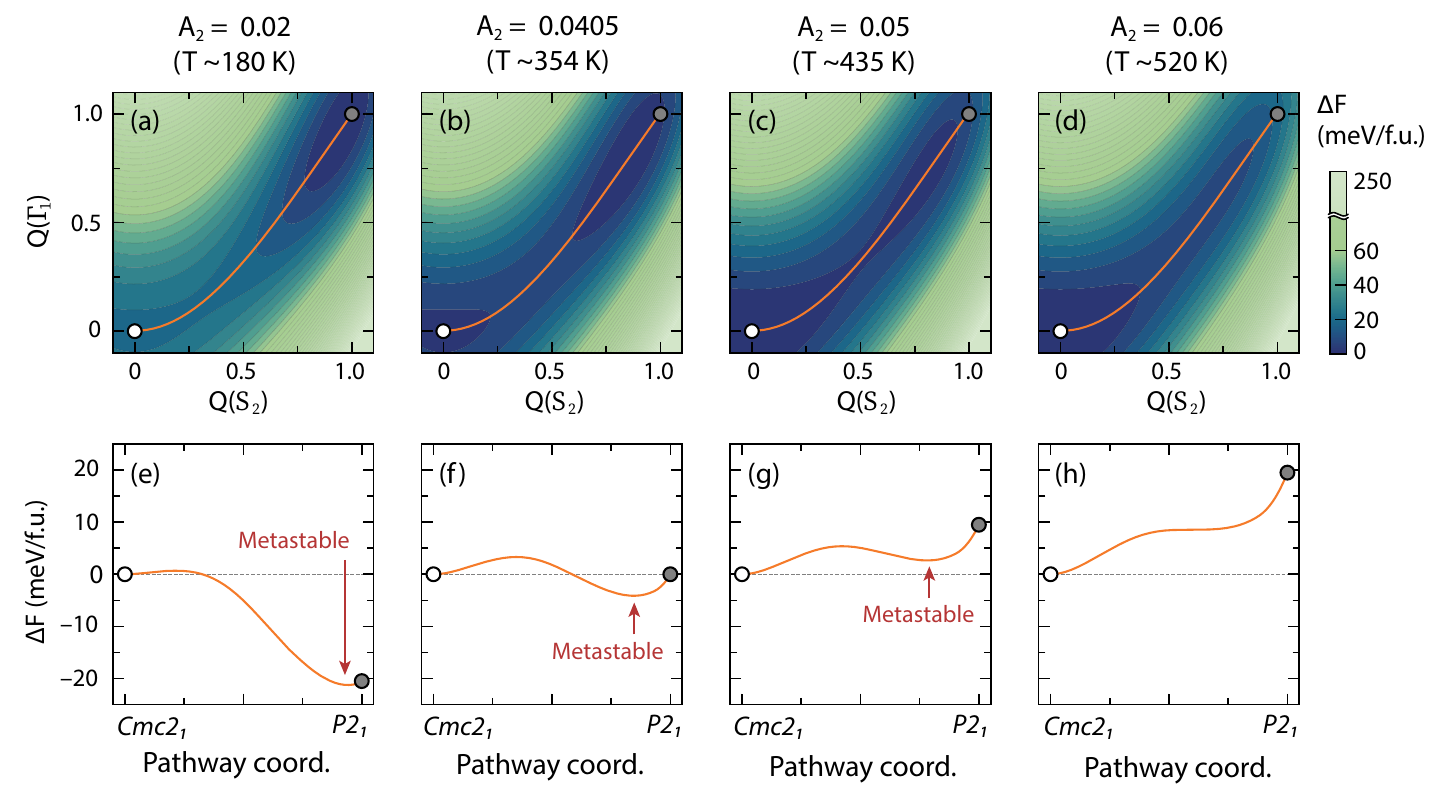}\vspace{-0.4em}
\caption{(a--d) Coupled $S_2$-$\Gamma_1$ Landau energy landscapes for $A_2=0.02$, 0.0405, 0.05, and 0.06, illustratively mapped to approximately 180, 354, 435, and 520 K using the linear calibration described in the text. White and grey circles denote $Cmc2_1$ and $P2_1$, and orange curves show the string-relaxed minimum-energy paths. (e--h) Corresponding energy profiles relative to $Cmc2_1$. The metastable $P2_1$ minimum disappears at $A_2=0.06$.}
\label{fig:temperature}
\vspace{1em}
\end{figure*}

The $S_2$ instability described above was obtained from the static 0 K energy landscape.
To examine phenomenologically how thermal stiffening could reshape this landscape, the harmonic coefficient $A_2$ is varied from 0.02 to 0.06 eV/f.u. The resulting energy surfaces and transition pathways are shown in \autoref{fig:temperature}. Distinct $Cmc2_1$ and $P2_1$ basins persist for $A_2=0.02$, 0.0405, and 0.05 eV/f.u., whereas the $P2_1$ basin becomes shallow and nearly disappears at $A_2=0.06$ eV/f.u.

Within a soft-mode description, thermal stabilization of the higher-symmetry $Cmc2_1$ phase is represented phenomenologically by
$A_2(T)=A_2(0)+\alpha T$.
The coefficient $\alpha$ is chosen such that $A_2=0.06$ eV/f.u. corresponds to 520 K, the experimental lower-temperature boundary of the $Cmc2_1$ phase \cite{Howieson/Morrison2020_Incommensurate,Howieson/Carpenter2021}.
This mapping provides a qualitative illustration of how the energy landscape may evolve with temperature, with all Landau coefficients other than $A_2$ held fixed.
The simultaneous presence of the two basins indicates bistability over a finite range of mode stiffness and is compatible with metastable phase competition and possible phase coexistence. The shift of the $P2_1$ minimum in mode-amplitude space further suggests that the $S_2$ modulation amplitude is temperature-sensitive within the assumed mapping.

Notably, varying a single coefficient does not constitute a complete finite-temperature free-energy model and therefore cannot predict a transition temperature.
The resulting temperature scale is illustrative rather than predictive because $B_2$ and the remaining Landau coefficients may also vary with temperature.
As a simplifying assumption, the fitted higher-order and coupling coefficients are held fixed while $A_2$ is varied.
The low-order expansion follows conventional Landau treatments of coupled order parameters \cite{Toledano1987landau,Salje/Carpenter2011,Young/Stroppa/Picozzi/Rondinelli2015}. Within this restricted model, the large negative coefficient of the $Q_{\Gamma_1}Q_{S_2}^2$ coupling maintains the $Cmc2_1$ and $S_2$-dominated $P2_1$ basins over a finite range of $A_2$.
The principal merit of this minimal treatment lies in isolating the effect of stiffening the primary $S_2$ coordinate while retaining the mode normalization used in the 0 K fit.
The estimated harmonic vibrational contributions increasingly favor $Cmc2_1$ with temperature but do not quantitatively reproduce the observed phase transitions \cite{Supp}. Although a more complete treatment of phonon renormalization may modify the quantitative value of $\alpha$, the present model’s qualitative conclusion remains that higher-order mode coupling can retain a metastable $P2_1$ minimum as thermal stiffening of $S_2$ increasingly favors $Cmc2_1$.

\begin{table*}[t]
\caption{\label{table:Lattice_parameters}
Experimental and calculated pseudo-orthorhombic lattice metrics of \LTO. Experimental values for the commensurate (C-o) and incommensurate (IC-o) phases are extrapolated to approximately 500 K from Ref.~\onlinecite{Howieson/Morrison2020_Incommensurate}, whereas the calculated values correspond to 0 K. The monoclinic cells follow the setting in \autoref{eq:transformation}. Here, $\Delta_\mathrm{exp}=\mathrm{IC\text{-}o}-\mathrm{C\text{-}o}$ and $\Delta_\mathrm{DFT}=P2_1-Cmc2_1$, with relative changes given in parentheses. The calculated lattice metrics of the $P2_1/c$ ground state are also included as a reference.
}
\centering
\renewcommand{\arraystretch}{1.2} 
\setlength{\tabcolsep}{10pt} 

\begin{tabular}{l c c c c c}
\hline \hline
Phase & $a_\text{o}$ (\AA) & $b_\text{o}$ (\AA) & $c_\text{o}$ (\AA) & $\gamma$ (\degree) & $V_\text{o}$ (\AA$^3$) \\
\hline 
\multicolumn{1}{l}{\Tstrut  \textbf{Experiment}} \\
C-o                     & 3.946                & 14.64               & 5.664                & 90   & 327.2 \\
IC-o                    & 3.943                & 14.72               & 5.646                & 90   & 327.7 \\
$\Delta_\mathrm{exp}$   & $-0.003$ ($-0.08\%$) & $+0.08$ ($+0.55\%$) & $-0.018$ ($-0.32\%$) & $0.0$ & $+0.5$ ($+0.15\%$) \\
\hline 
\multicolumn{1}{l}{\Tstrut  \textbf{DFT calculations}} \\
$Cmc2_1$                & 3.944                & 14.67               & 5.600                & 90.0 & 324.0 \\
$P2_1$                  & 3.898                & 15.00               & 5.545                & 89.9 & 324.2 \\
$\Delta_\mathrm{DFT}$   & $-0.046$ ($-1.17\%$) & $+0.33$ ($+2.25\%$) & $-0.055$ ($-0.98\%$) & $-0.1$ & $+0.2$ ($+0.06\%$) \Bstrut \\
\hline 
\multicolumn{1}{l}{\Tstrut  \textbf{DFT ground state}} \\
$P2_1/c$                & 3.893                & 15.01               & 5.530                & 86.3 & 322.4 \\
\hline \hline
\end{tabular}
\end{table*}

\subsection{\boldmath Connection between the commensurate $P2_1$ reference and the incommensurate phase}

Here, the calculated $P2_1$ phase serves as a commensurate reference for examining local structural motifs relevant to IC-o, although the two phases differ in translational periodicity.
Neutron diffraction describes IC-o using the basic space group $Cmc2_1$ and a modulation vector $q\simeq(0.456,0,0)$ \cite{Howieson/Morrison2020_Incommensurate,Howieson/Carpenter2021}, whereas in situ electron diffraction gives $q\simeq(0.4669,0,0)$ and the superspace group $Cmc2_1(\alpha00)0s0$ \cite{Wu/Howieson/Ma2026_In_situ}.
The calculations, by contrast, sample the commensurate $S$-point instability at $q=(1/2,0,0)_{\mathrm{prim}}$, equivalent to $(1/2,1/2,0)$ in the conventional orthorhombic reciprocal basis, producing a metastable $P2_1$ structure.
Accordingly, the calculated $P2_1$ phase is treated as a commensurate reference for IC-o rather than as an explicit structural assignment of the IC-o phase.

With this distinction established, the lattice response along the phase transitions provides the first basis for comparison. To compare the lattice parameters of the monoclinic and orthorhombic cells in a common setting,
the monoclinic axes are transformed into a pseudo-orthorhombic convention aligned with the conventional orthorhombic axes \cite{Wu/Howieson/Ma2026_In_situ}:
\begin{equation}
\begin{pmatrix}
a_{o} \\ b_{o} \\c_{o}
\end{pmatrix}
=
\begin{pmatrix}
0 & 0 & \nicefrac{1}{2} \\ 2 & 0 & \nicefrac{1}{2} \\ 0 & 1 & 0 
\end{pmatrix}
\begin{pmatrix}
a_m \\ b_m \\ c_m
\end{pmatrix}.
\label{eq:transformation}
\end{equation}
The transformation permits direct comparison of the lattice metrics summarized in \autoref{table:Lattice_parameters}. 

Here, relative changes are more informative than the absolute lattice constants, because the calculations refer to 0 K while the experimental values are extrapolated to approximately 500 K.
Across the experimental C-o-to-IC-o transition, $a_\mathrm{o}$ and $c_\mathrm{o}$ decrease slightly, while $b_\mathrm{o}$ and the cell volume increase \cite{Howieson/Morrison2020_Incommensurate}.
Similarly, the calculated $Cmc2_1\rightarrow P2_1$ transformation produces changes with the same signs, although the calculated magnitudes are larger.
In addition, the pseudo-orthorhombic $P2_1$ cell retains $\gamma=89.9^\circ$, close to $\gamma=90^\circ$ for the conventional orthorhombic $Cmc2_1$ cell.
For comparison, the pseudo-orthorhombic lattice parameters of the $P2_1/c$ ground state are also considered. Relative to $Cmc2_1$, the $P2_1/c$ values show the same axial trends as the $P2_1$ values, but the $P2_1/c$ cell has a smaller volume and a substantially larger angular deviation, with $\gamma=86.3^\circ$.
Thus, among the calculated monoclinic structures, $P2_1$ more closely reproduces the near-orthorhombic average lattice response observed across the C-o-to-IC-o transition.

\begin{figure}[h]
\centering
\includegraphics[width=0.85\columnwidth]{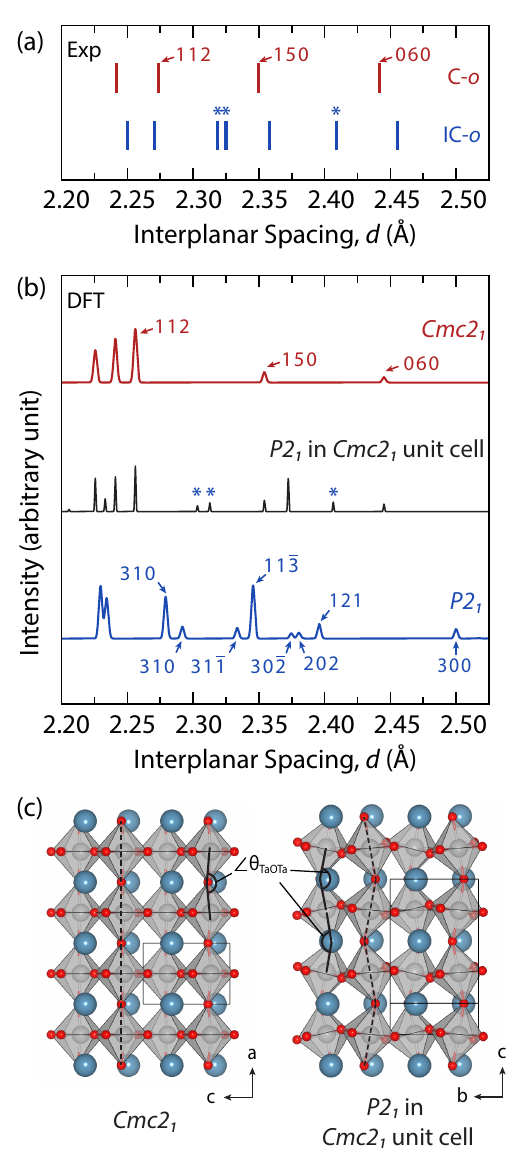}
\caption{(a) Experimental neutron diffraction spectra of the commensurate orthorhombic phase at 520 K and the incommensurate orthorhombic phase at 480 K from Ref.~\onlinecite{Howieson/Morrison2020_Incommensurate}. Asterisks mark reflections associated with the incommensurate modulation. (b) Simulated patterns for $Cmc2_1$, $P2_1$ constrained to the $Cmc2_1$ lattice metrics, and fully relaxed $P2_1$. Curves are vertically offset, and intensities are given in arbitrary units.
(c) Structures of $Cmc2_1$ (left) and $P2_1$ constrained to the $Cmc2_1$ lattice metrics (right). The indicated Ta–O–Ta bond angle ($\angle_{\text{TaOTa}}$) has a single value in $Cmc2_1$ and splits into two distinct values in $P2_1$. The angles for $P2_1$ structure in $Cmc2_1$ unit cell are 156.2\degree and 161.7\degree.
}
\label{fig:Neutron}
\end{figure}

In addition to the compatible lattice response, the atomic displacement patterns are compared using the experimental and simulated neutron-diffraction patterns shown in \autoref{fig:Neutron}.
Experimentally, the C-o and IC-o patterns share the principal $(112)$, $(150)$, and $(060)$ reflections of the basic orthorhombic cell, whereas additional modulation-associated features near $d=2.32$--2.33~\AA\ and $d=2.41$~\AA\ distinguish IC-o \cite{Howieson/Morrison2020_Incommensurate}.
Consistent with experiment, the simulated $Cmc2_1$ pattern reproduces the principal C-o reflections, with small shifts in interplanar spacing arising from differences between the calculated and experimental lattice metrics.
By contrast, the fully relaxed $P2_1$ pattern differs substantially from the IC-o pattern.
One obvious reason for the mismatch is the difference in unit-cell shape. For example, the $P2_1$ $(300)$ reflection can be related to the layer periodicity represented by the $Cmc2_1$ $(060)$ reflection, but the different lattice parameters shift the two reflections to different interplanar spacings.

Because the mismatch contains contributions from both the displacement pattern and the lattice strain, a constrained $P2_1$ pattern was calculated using the $Cmc2_1$ unit-cell shape [black curve in \autoref{fig:Neutron}(b)].
Note that the orthorhombic constraint was expressed in the corresponding monoclinic setting using \autoref{eq:transformation}.
Under the $Cmc2_1$ lattice constraint, the $P2_1$ $(300)$ reflection shifts toward the experimental $(060)$ position, while the $(31\bar{1})$, $(310)$, and $(121)$ reflections occur near the modulation-associated IC-o features marked by asterisks.
Nevertheless, additional commensurate reflections remain absent from the experimental pattern, consistent with the different modulation vectors.
For example, the absence of the $P2_1$ $(11\bar{3})$ reflection in the IC-o pattern further illustrates the effect of incommensurate modulation. The large $c$ index probes the monoclinic direction corresponding to the orthorhombic $a$ axis, along which the IC-o modulation propagates.
The partial correspondence therefore supports a related $S_2$-dominated displacement character, while $Cmc2_1(\alpha00)0s0$ remains the appropriate crystallographic description of IC-o \cite{Wu/Howieson/Ma2026_In_situ}.

The differences between the diffraction patterns are consistent with related octahedral-rotation patterns whose amplitudes vary over a longer range in IC-o.
Unlike $P2_1/c$ and $Cmc2_1$, the calculated $P2_1$ structure exhibits two distinct Ta--O--Ta angles ($\angle_{\text{TaOTa}}$) for the corresponding linkages [\autoref{fig:Neutron}(c)]. This splitting resembles that reported for IC-o by Wu et al.  \cite{Wu/Howieson/Ma2026_In_situ}, further supporting the local structural similarity between IC-o and $P2_1$ phases.
Moreover, the reflection shifts produced by the $Cmc2_1$ lattice constraint suggest that coherency strain could modify the diffraction signatures of locally stabilized $P2_1$-like regions within a $Cmc2_1$ matrix. Such local configurations are compatible with the competing $Cmc2_1$ and $P2_1$ basins in \autoref{fig:temperature}. Although the static energy surfaces omit explicit temperature effects, the metastable $q=1/2$ basin provides a commensurate reference for identifying strain-stabilized or locally commensurate configurations in the finite-temperature transition landscape.

Beyond the direct structural comparison, the commensurate reference provides a concrete starting point for investigating possible Devil's-staircase behavior.
In this context, the measured $q(T)$ dependence motivates consideration of temperature-dependent lock-in states with $q=m/n$ \cite{Howieson/Morrison2020_Incommensurate,Scott1982,Scott2011,Cox1979,St-gregoire/Stigenberger1984}.
Previous studies of the IC-o phase have focused mainly on structural modulation near the high-temperature phase boundary. Establishing $P2_1$ as a commensurate reference enables future studies to examine electronic contributions to modulation stability over a broader temperature range \cite{Mitchell1995}.
Systematic comparison among rational approximants could clarify how lattice strain, modulation amplitude, and the energetic cost of phase slips or discommensurations govern possible lock-in behavior.

\subsection{Chemical trends in Carpy-Galy phase competition}

\begin{figure*}[t]
\centering
\includegraphics[width=0.75\textwidth]{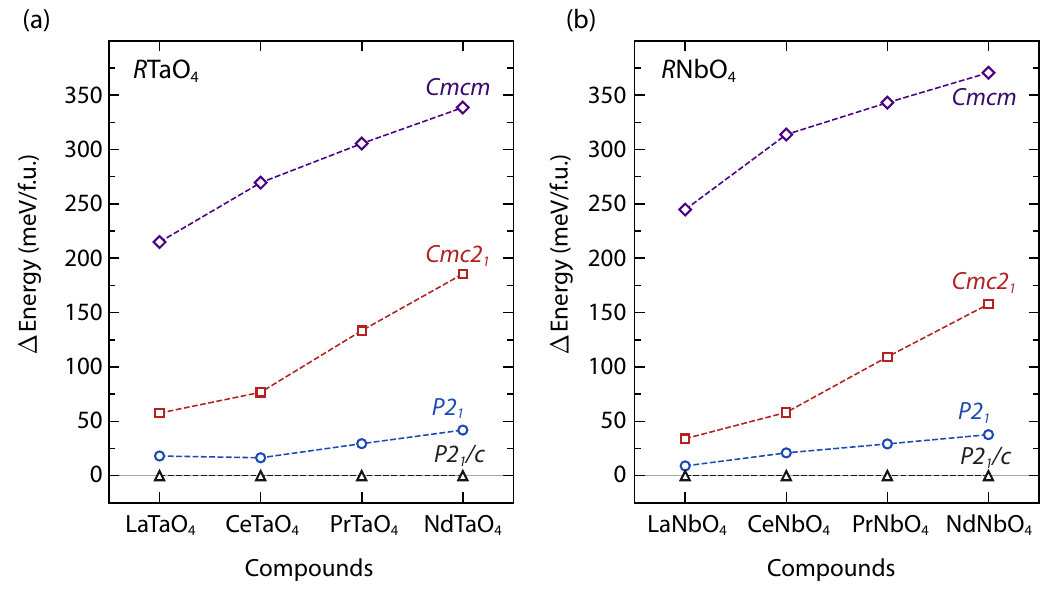}
\caption{Calculated 0 K energies of the $P2_1/c$, $P2_1$, $Cmc2_1$, and $Cmcm$ polymorphs across the (a) $R$TaO$_4$ and (b) $R$NbO$_4$ series ($R=$ La, Ce, Pr, and Nd). Energies are given relative to $P2_1/c$ for each composition.}
\label{fig:VariousCompounds}
\end{figure*}

Chemical tuning of Carpy-Galy phase competition must first account for the narrow stability range of the $n=2$ topology.
The $n=2$ Carpy–Galy structure has been reported for LaTaO$_4$, CeTaO$_4$, and PrTaO$_4$, illustrating its restricted compositional range \cite{Cava1981,Vullum/Grande2008,Howieson/Morrison2020_Incommensurate,Cui/Tian2024}.
The limited compositional range is attributed to strong competition from scheelite and fergusonite $RB$O$_4$ compounds, which exhibit edge-connected octahedra or isolated tetrahedra rather than the perovskite-derived octahedral slabs of the Carpy-Galy structure \cite{Dias2017,Vullum/Grande2008,Yang2018,Banerjee2025,Garg2024,Wolten1967,Errandonea2019}.
Among the surveyed compositions, only LaTaO$_4$ has a Carpy-Galy $P2_1/c$ structure that is lower in energy than both the fergusonite and scheelite polymorphs, whereas the other $RB$O$_4$ compounds energetically favor one of these two polymorphs \cite{Supp}.
Thus, the substitution trends discussed here refer to relative energetics within the constrained Carpy-Galy polymorph manifold.

Nevertheless, partial substitution has proved possible while preserving the LaTaO$_4$ Carpy-Galy framework. For example, approximately 20\% Nb can replace Ta in LaNb$_{1-x}$Ta$_x$O$_4$ for $x>0.8$, while approximately 10\% Nd substitution has been realized on the La site \cite{Cordrey/Lightfoot2015,Vullum/Grande2008,Howieson/Morrison2020_Incommensurate}. These experimentally accessible solid solutions motivate an examination of how $R$- and $B$-site chemistry modifies competition among the Carpy-Galy polymorphs \cite{Shin/Galli2023,NunezValdez/Spaldin2019_Origin,Han/Shin/Galli2026,Shin/Leighton/Galli2025}.

Against this chemical background, \autoref{fig:VariousCompounds} compares the relative energies within the Carpy-Galy topology. 
The $P2_1/c$ structure remains the lowest-energy Carpy-Galy polymorph for every composition examined. 
As the rare-earth radius decreases from La to Nd, the energies of $Cmc2_1$ and $Cmcm$ generally rise relative to $P2_1/c$, whereas the increase for $P2_1$ is more modest. 
The differential response is consistent with a growing elastic cost for accommodating smaller $R$-site cations in less favorable coordination environments \cite{NunezValdez/Spaldin2019_Origin,Shin/Rondinelli2024,Shin/Rondinelli2021,Shin/Rondinelli2020,Shin/Shin2025_JACS}.
The $P2_1$ phase deviates from a strictly monotonic trend with ionic radius: the energy difference between $P2_1$ and $P2_1/c$ decreases slightly from LaTaO$_4$ to CeTaO$_4$ before increasing across PrTaO$_4$ and NdTaO$_4$.

To characterize the possible competition between the C-o and IC-o phases, the most relevant quantity is the energy difference between the $P2_1$ and $Cmc2_1$ phases, defined as
\begin{equation}
\Delta E_{Cmc2_1-P2_1} = E(Cmc2_1)-E(P2_1).
\end{equation}
Because the calculated $P2_1$ structure shares local structural motifs with IC-o, $\Delta E_{Cmc2_1-P2_1}$ provides a qualitative 0 K descriptor for examining chemical trends relevant to this phase competition.
Across the $R$TaO$_4$ series, $\Delta E_{Cmc2_1-P2_1}$ increases as the rare-earth radius decreases, indicating greater stabilization of $P2_1$ relative to $Cmc2_1$. 
Indeed, partial Nd substitution raises the transition temperature \cite{Howieson/Morrison2020_Incommensurate}, paralleling the calculated increase in $\Delta E_{Cmc2_1-P2_1}$. Extending the same rare-earth trend therefore motivates the testable hypothesis that partial substitution of Ce, Pr, or Nd for La may raise the transition temperature.

By contrast, changing the $B$-site from Ta to Nb produces the opposite trend. Despite the similar tabulated ionic radii of Ta and Nb \cite{Shannon1976}, $\Delta E_{Cmc2_1-P2_1}$ is smaller for the $R$NbO$_4$ compositions than for the corresponding $R$TaO$_4$ compositions (\autoref{fig:VariousCompounds}).
Thus, Nb stabilizes $Cmc2_1$ relative to the commensurate $P2_1$ reference.
The calculated trend is consistent with the reported decrease in the lower-temperature boundary of the $Cmc2_1$ phase upon Nb substitution \cite{Vullum/Grande2008}. 
However, the reported boundary was determined by comparison with the nonpolar $P2_1/c$ phase, leaving the IC-o phase unresolved for Nb-substituted compositions.
Consequently, the influence of Nb on the C-o-to-IC-o boundary remains open for experimental investigation.

\section{Conclusion}
First-principles calculations identify the polar monoclinic $P2_1$ structure of Carpy-Galy \LTO as a metastable $q=1/2$ minimum at 0 K, with an energy between those of $P2_1/c$ and $Cmc2_1$.
As a common subgroup of $P2_1/c$ and $Cmc2_1$, the $P2_1$ phase provides a metastable intermediate along a symmetry-connected route between the established antipolar and polar structures. Coupling between the structural order parameter and an isosymmetric lattice mode provides substantial stabilization, and the resulting energy surfaces reveal a finite transformation barrier and competing local minima.
Although the calculated and experimental structures differ in translational periodicity,
lattice and diffraction comparisons are compatible with a shared $S_2$-dominated displacement character, enabling $P2_1$ to serve as a commensurate reference for examining possible Devil's-staircase behavior.
Calculated trends across selected $R$TaO$_4$ and $R$NbO$_4$ compositions further show how substitution shifts the relative stabilities of the Carpy-Galy polymorphs, suggesting chemical routes to tune polar-antipolar competition within the stability range of LaTaO$_4$.

\begin{acknowledgments}
The present research was supported by the research fund of Dankook University in 2025.

\end{acknowledgments}

\hyphenpenalty=50
\exhyphenpenalty=50
\bibliography{StructureMap}

\end{document}


\noindent
{\bf Supplementary Material for}\\
\vspace{-0.3 cm}

\noindent{\bf \boldmath Metastable polar order and phase competition in Carpy-Galy LaTaO$_4$}
\vspace{0.5cm}

\noindent
Yongjin Shin$^{1,*}$\vspace{0.3cm}\\
\renewcommand{\baselinestretch}{0.90}\normalsize
$^1$\textit{\footnotesize Department of Semiconductor Convergence Engineering, Dankook University, Yongin-si 16890, Gyeonggi-do, South Korea}\\
$^*${\small Email: yongjin.shin@dankook.ac.kr}\\

\vspace{0.75cm}

\def\degree{$^\circ$\xspace}
\newcommand\Tstrut{\rule{0pt}{2.6ex}}         
\newcommand\Bstrut{\rule[-0.9ex]{0pt}{0pt}}   
\newcommand\RAFO{$R_{1/3}A_{2/3}$FeO$_{2.67}$\xspace}
\newcommand\NCFO{Nd$_{1/3}$Ca$_{2/3}$FeO$_{2.67}$\xspace}
\newcommand\LSFO{La$_{1/3}$Sr$_{2/3}$FeO$_{2.67}$\xspace}
\newcommand\SFO{SrFeO$_{2.5}$\xspace}
\newcommand\CFO{CaFeO$_{2.5}$\xspace}
\newcommand\BFO{BaFeO$_{2.5}$\xspace}
\newcommand\AFO{$A$FeO$_{2.5}$\xspace}

\renewcommand{\baselinestretch}{0.95}\normalsize
{
\hypersetup{linkcolor=black}
\tableofcontents
}
\renewcommand{\baselinestretch}{1.2}\normalsize


\newpage

\section[Crystallographic data for LaTaO$_4$ variants in the Carpy-Galy structure]{\boldmath Crystallographic data for LaTaO$_4$ variants in the Carpy-Galy structure}
\label{sec:crystallographic}

\begin{table}[htbp]
\caption{\label{table:Cmcm} 
Crystallographic data for \LTO in the $Cmcm$ space group (No. 63), obtained by total-energy optimization with the PBEsol functional. Lattice parameters are $a$ = 3.9659 \AA, $b$ = 15.3528 \AA, $c$ = 5.5694 \AA, $\alpha$ = $\beta$ = $\gamma$ = 90\degree, and the unit-cell volume ($V$) is 339.11 \AA$^3$. 
The Ta--O--Ta angle ($\angle_{\text{TaOTa}}$) is 172.53\degree.
}
\centering
\vspace{0.5em}
\setlength{\tabcolsep}{12pt} 
\begin{tabular}{l|c|ccc}
\hline\hline
Atom & Site & x & y & z \\
\hline\Tstrut
La1  & 4c  &  0.00000 &  0.18553  & 0.25000  \\
Ta1  & 4c  &  0.00000 &  0.41200  & 0.25000  \\
O1   & 8f  &  0.00000 &  0.31555  & 0.49502 \\
O2   & 4b  &  0.00000 &  0.50000  & 0.00000  \\
O3   & 4c  &  0.00000 &  -0.09643  & 0.25000  \\
\hline\hline
\end{tabular}
\end{table}
%

\begin{table}[htbp]
\caption{\label{table:Cmc21} 
Crystallographic data for \LTO in the $Cmc2_1$ space group (No. 36), obtained by total-energy optimization with the PBEsol functional. Lattice parameters are $a$ = 3.9442 \AA, $b$ = 14.6708 \AA, $c$ = 5.6000 \AA, $\alpha$ = $\beta$ = $\gamma$ = 90\degree, and the unit-cell volume ($V$) is 324.05 \AA$^3$.
The Ta--O--Ta angle ($\angle_{\text{TaOTa}}$) is 169.19\degree.
}
\centering
\vspace{0.5em}
\setlength{\tabcolsep}{12pt} 
\begin{tabular}{l|c|ccc}
\hline\hline
Atom & Site & x & y & z \\
\hline\Tstrut
La1  & 4a  &  0.00000 &  0.17282 & 0.16932  \\
Ta1  & 4a  &  0.00000 &  0.41441 & 0.21873  \\
O1   & 4a  &  0.00000 &  0.30023 & 0.42650  \\
O2   & 4a  &  0.00000 &  0.33507 & -0.05130 \\
O3   & 4a  &  0.00000 &  0.46880 & 0.56176  \\
O4   & 4a  &  0.00000 &  -0.08769& 0.25160  \\
\hline\hline
\end{tabular}
\end{table}

\begin{table}[htbp]
\caption{\label{table:P21c} 
Crystallographic data for \LTO in the $P2_1/c$ space group (No. 14), obtained by total-energy optimization with the PBEsol functional. Lattice parameters are $a$ = 7.6275 \AA, $b$ = 5.5304 \AA, $c$ = 7.7862 \AA, $\alpha$ = $\gamma$ = 90\degree, $\beta$ = 101.02\degree, and the unit-cell volume ($V$) is 322.39 \AA$^3$.
The Ta--O--Ta angle ($\angle_{\text{TaOTa}}$) is 151.85\degree.
}
\centering
\vspace{0.5em}
\setlength{\tabcolsep}{12pt} 
\begin{tabular}{l|c|ccc}
\hline\hline
Atom & Site & x & y & z \\
\hline\Tstrut
La1  & 4e  & 0.3465 & 0.2729 & 0.5980 \\
Ta1  & 4e  & 0.1672 & 0.7665 & 0.8011 \\
O1   & 4e  & 0.1677 & 0.6617 & 0.5531 \\
O2   & 4e  & 0.0545 & 0.0854 & 0.7062 \\
O3   & 4e  & 0.3824 & -0.0138 &0.8327 \\
O4   & 4e  & 0.3375 & 0.5097 & 0.8641 \\
\hline\hline
\end{tabular}
\end{table}

\begin{table}[htbp]
\caption{\label{table:P21} 
Crystallographic data for \LTO in the $P2_1$ space group (No. 4), obtained by total-energy optimization with the PBEsol functional. Lattice parameters are $a$ = 7.7446 \AA, $b$ = 5.5452 \AA, $c$ = 7.7965 \AA, $\alpha$ = $\gamma$ = 90\degree, $\beta$ = 104.44\degree, and the unit-cell volume ($V$) is 324.25 \AA$^3$.
The Ta--O--Ta angles ($\angle_{\text{TaOTa}}$) are 153.32\degree and 156.63\degree.
}
\centering
\vspace{0.5em}
\setlength{\tabcolsep}{12pt} 
\begin{tabular}{l|c|ccc}
\hline\hline
Atom & Site & x & y & z \\
\hline\Tstrut
La1 &   2a &  0.3196  &  0.2178  &  0.8365  \\
La2 &   2a &  0.3696  &  0.1684  &  0.3411  \\
Ta1 &   2a &  0.8315  &  0.2220  & -0.0464 \\
Ta2 &   2a &  0.8326  &  0.2251  &  0.4553  \\
O1  &   2a &  0.6057  &  0.4256  & -0.0808 \\
O2  &   2a &  0.6187  &  0.4367  &  0.3850  \\
O3  &   2a &  0.6842  & -0.0555  &  0.8989  \\
O4  &   2a &  0.6616  & -0.0411  &  0.4313  \\
O5  &   2a & -0.0569  &  0.5538  &  0.0334  \\
O6  &   2a & -0.0476  &  0.5434  &  0.4363  \\
O7  &   2a &  0.8408  &  0.1516  &  0.2091  \\
O8  &   2a &  0.8207  &  0.3052  &  0.7037  \\
\hline\hline
\end{tabular}
\vspace{1em} 
\end{table}

\newpage

\section[Relationships among the calculated $P2_1$, experimental $m'$, and IC-o structures]{\boldmath Relationships among the calculated $P2_1$, experimental $m'$, and IC-o structures}

The calculated metastable structure and the transient $m'$ phase recently identified by in situ electron diffraction share the $P2_1$ space-group label but have different translational symmetries. The experimental $m'$ phase has a short repeat of approximately 3.97~\AA\ along the $c$ direction, whereas the corresponding repeat in the calculated $P2_1$ cell is about twice as long. The shared space-group label alone therefore does not justify assigning the calculated structure to the experimental $m'$ phase. \autoref{table:structure_comparison} summarizes the distinct crystallographic relationships and physical roles.

\begin{table*}[htbp]
\caption{\label{table:structure_comparison}
Comparison of the transient experimental $m'$ phase at 380 K, the calculated commensurate $P2_1$ state, and the experimental IC-o phase at 400 K. The experimental lattice parameters are from Ref.~\onlinecite{Wu/Howieson/Ma2026_In_situ}. The reported modulation range combines $q=0.456$ at 483 K from Ref.~\onlinecite{Howieson/Morrison2020_Incommensurate} and $q=0.4669$ at 400 K from Ref.~\onlinecite{Wu/Howieson/Ma2026_In_situ}. Lattice parameters are in \AA. The subscripts $\mathrm{prim}$ and $\mathrm{conv}$ distinguish the primitive and conventional orthorhombic reciprocal bases, respectively, with $(1/2,0,0)_{\mathrm{prim}}=(1/2,1/2,0)_{\mathrm{conv}}$.
}
\centering
\small
\setlength{\tabcolsep}{5pt}
\begin{tabular}{l c c c c c c}
\hline\hline
Structure & Symmetry & $a$ & $b$ & $c$ & $\beta$ & $q$ \\
\hline\Tstrut
Experimental $m'$ (380 K) & $P2_1$ & 7.78 & 5.65 & 3.97 & 100.94$^\circ$ & -- \\
Calculated state & $P2_1$ & 7.7446 & 5.5452 & 7.7965 & 104.44$^\circ$ & $(1/2,0,0)_{\mathrm{prim}}$ \\
Experimental IC-o (400 K) & $Cmc2_1(\alpha00)0s0$ & 3.95 & 14.58 & 5.69 & 90$^\circ$ & $(0.456\text{--}0.4669,0,0)_{\mathrm{conv}}$ \\
\hline\hline
\end{tabular}
\end{table*}

\clearpage

\section[Symmetry breaking from $Cmcm$ to $Cmc2_1$ and $P2_1/c$ structures]{\boldmath Symmetry breaking from $Cmcm$ to $Cmc2_1$ and $P2_1/c$ structures}

The symmetry-breaking pathways from the high-symmetry $Cmcm$ phase to the lower-symmetry $Cmc2_1$ and $P2_1/c$ polymorphs are governed by distinct phonon modes. The pathway to polar $Cmc2_1$ is proper ferroelectric and is driven by the $\Gamma_2^-$ symmetry-breaking mode as the sole primary order parameter. Structural mode decomposition shows that the isosymmetric $\Gamma_1^+$ mode accounts for 4.5\% of the total distortion and primarily describes outward La displacements from the perovskite slabs, whereas the primary $\Gamma_2^-$ mode contributes 95.5\%. Although the isolated $\Gamma_1^+$ mode raises the energy monotonically, strong coupling between $\Gamma_1^+$ and the primary $\Gamma_2^-$ instability deepens the $Cmc2_1$ energy minimum, consistent with geometrically induced ferroelectricity.

\begin{figure*}[htbp]
\includegraphics[width=0.80\textwidth]{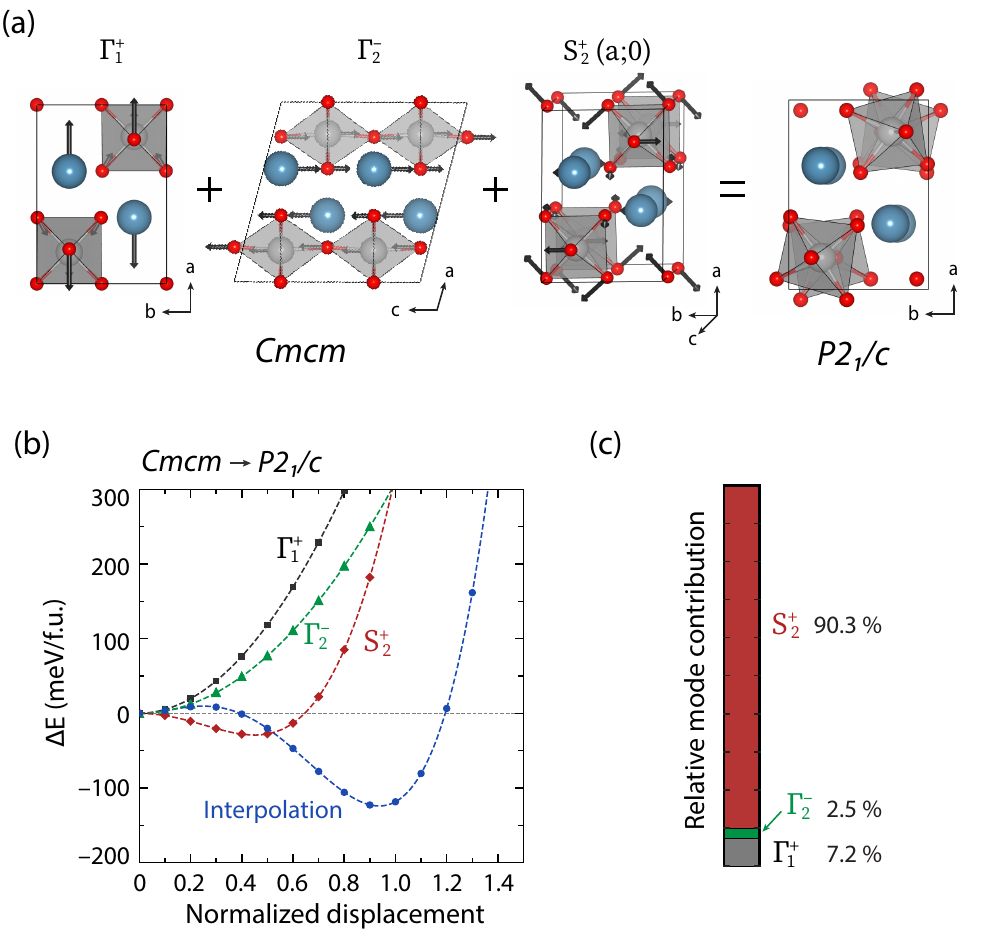}
\caption{\label{fig:S_Cmcm_P21c} 
(a) Displacement patterns for the principal $\Gamma_1^+$, $\Gamma_2^+$, and $S_2^+$ modes connecting $Cmcm$ and $P2_1/c$. (b) Energies of the isolated modes and the geometrically constrained linear interpolation as functions of normalized displacement amplitude, referenced to $Cmcm$. The interpolation maximum is path dependent and does not represent the minimum barrier. (c) Relative mode contributions to the $P2_1/c$ structure, dominated by $S_2^+$ (90.3\%).
}
\vspace{1.5cm}
\end{figure*}

By contrast, the pathway to ground-state $P2_1/c$ requires a multidimensional description. A linear structural interpolation from $Cmcm$ exhibits an apparent energy maximum, consistent with previous theoretical work \cite{Liu/Chen2016_Topological}. Mode decomposition and the corresponding Landau energy surface show that the maximum is imposed by the geometrically constrained one-dimensional interpolation. Within the sampled multidimensional mode space, relaxation initiated along $S_2^+$ follows a monotonically decreasing trajectory from $Cmcm$ to $P2_1/c$. The interpolation maximum in \autoref{fig:S_Cmcm_P21c}(b) is therefore path dependent and should not be interpreted as the minimum barrier for the transformation. This result is distinct from the SS-NEB barrier between $P2_1/c$ and metastable $P2_1$ reported in the main text.

\begin{figure*}[htbp]
\includegraphics[width=0.95\textwidth]{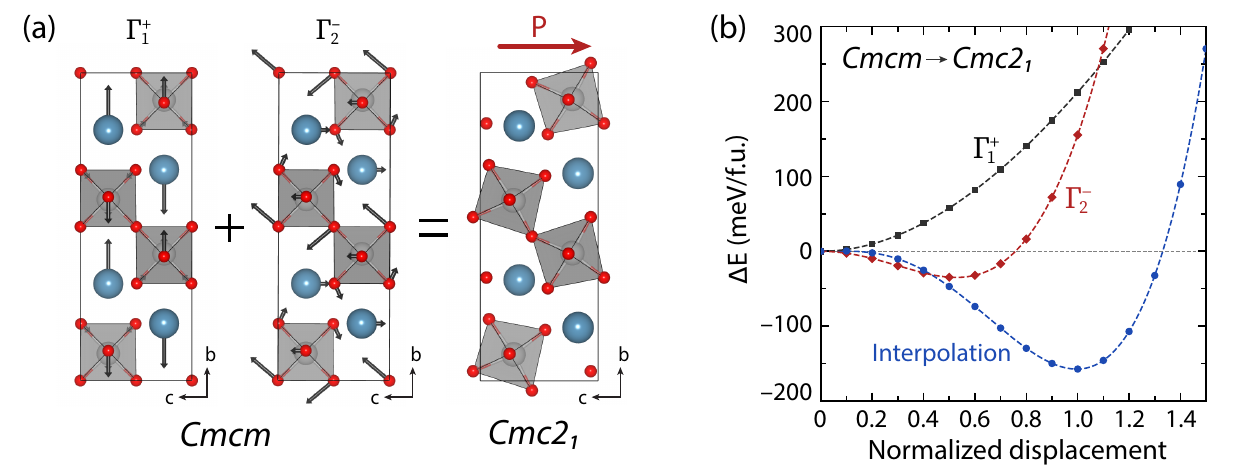}
\caption{\label{fig:S_Cmcm_Cmc21} 
(a) Structural representations of the $\Gamma_1^+$ and $\Gamma_2^-$ displacement vectors connecting nonpolar $Cmcm$ and polar $Cmc2_1$. (b) Energies of the isolated modes and the combined linear interpolation as functions of normalized displacement amplitude, referenced to $Cmcm$. The $\Gamma_1^+$ and $\Gamma_2^-$ modes account for 4.5\% and 95.5\% of the total structural distortion, respectively.
}
\vspace{1.5cm}
\end{figure*}

\clearpage
\section[Phonon dispersions of Carpy-Galy LaTaO$_4$]{\boldmath Phonon dispersions of Carpy-Galy LaTaO$_4$}

\begin{figure*}[htbp]
\includegraphics[width=0.99\textwidth]{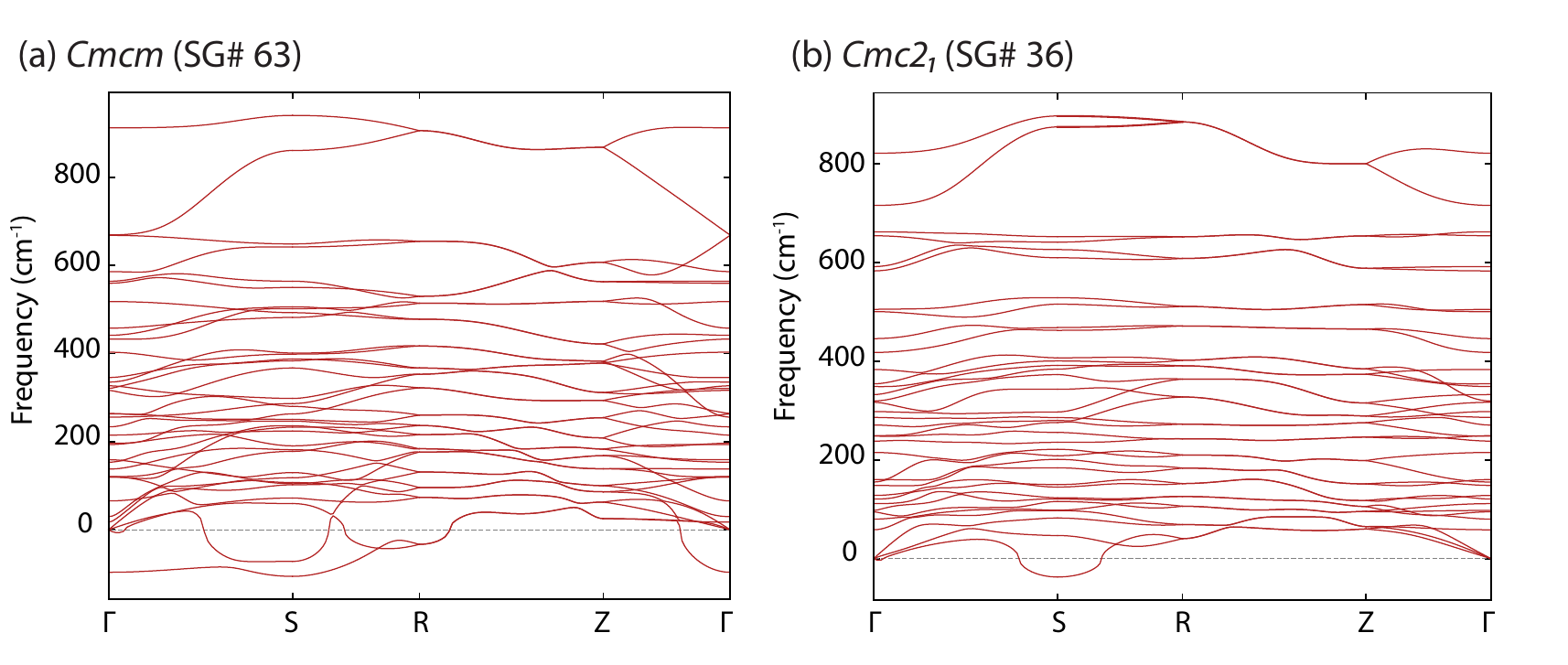}
\vspace{-1em}
\caption{\label{fig:S_PhononOrtho} 
(a) Phonon dispersion of the parent $Cmcm$ structure (No. 63), showing extensive dynamical instabilities. (b) Phonon dispersion of polar $Cmc2_1$ (No. 36), with a distinct imaginary frequency at the $S$ point. In the reciprocal-lattice convention of the base-centered orthorhombic cell, $S$ corresponds to $q_S=(1/2,0,0)$, and the associated $S_2$ distortion leads toward the commensurate $P2_1$ structure.
}
\vspace{0.2cm}
\end{figure*}

\begin{figure*}[htbp]
\includegraphics[width=0.55\textwidth]{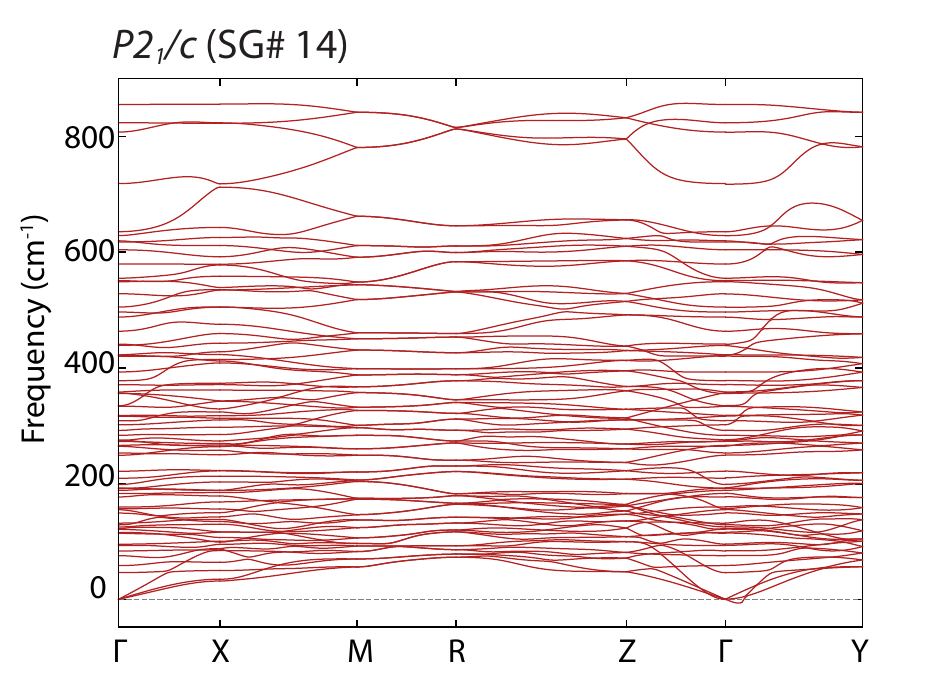}
\vspace{-1em}
\caption{\label{fig:S_PhononP21c} 
Phonon dispersion of the LaTaO$_4$ $P2_1/c$ phase (No. 14). The absence of imaginary modes along the sampled high-symmetry paths confirms dynamical stability at 0 K. Together with the lowest total energy among the examined polymorphs, the result identifies antipolar $P2_1/c$ as the global minimum on the calculated energy landscape.
}
\vspace{0.2cm}
\end{figure*}

The phonon dispersions of the $Cmcm$, $Cmc2_1$, and $P2_1/c$ polymorphs are shown in \autoref{fig:S_PhononOrtho} and \autoref{fig:S_PhononP21c}. The high-symmetry $Cmcm$ structure exhibits several imaginary branches and is therefore dynamically unstable at 0~K. Condensation of the zone-center polar instability produces $Cmc2_1$, which retains a distinct $S$-point instability associated with the $S_2$ distortion toward the commensurate $P2_1$ state. In contrast, the $P2_1/c$ ground state has no imaginary frequencies along the sampled high-symmetry paths, consistent with dynamical stability.

\begin{figure*}[htbp]
\includegraphics[width=0.99\textwidth]{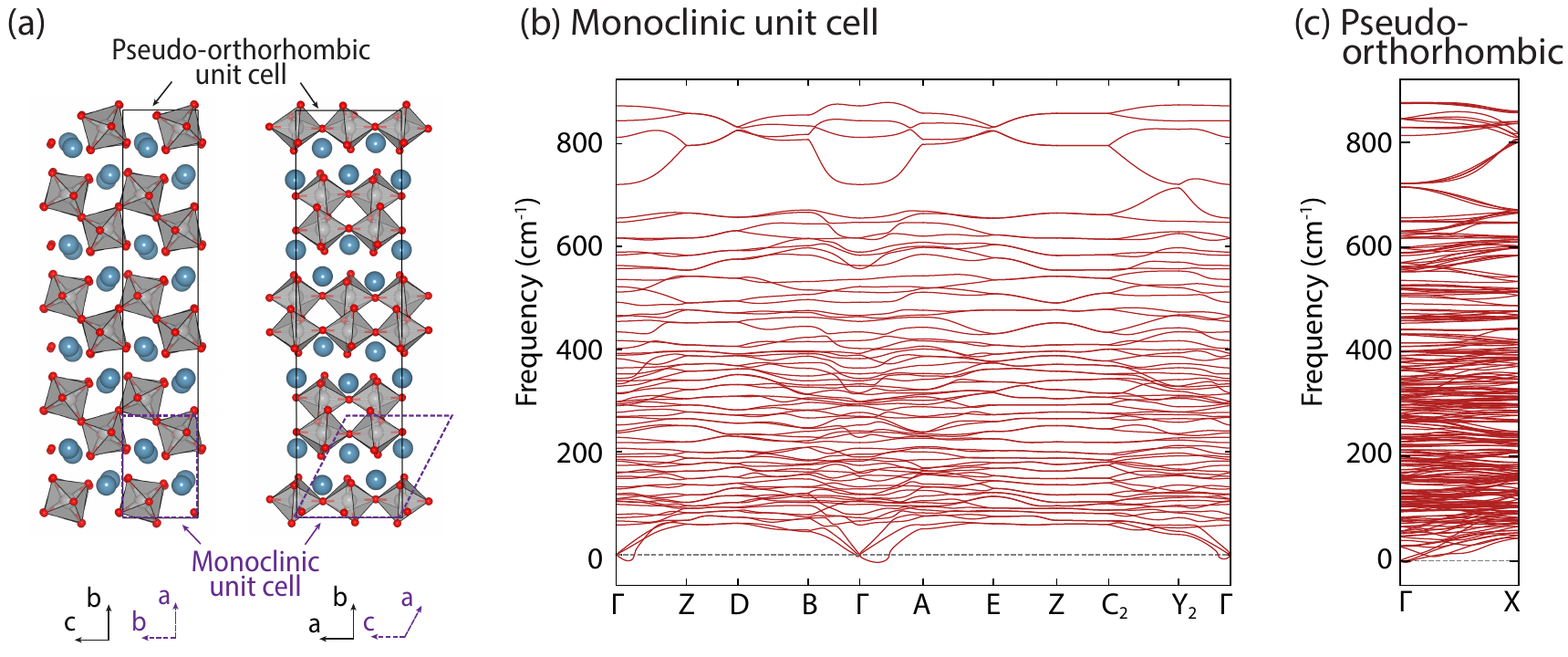}
\caption{\label{fig:S_PhononP21} 
Phonon dispersion mapping and dynamical stability of the calculated $P2_1$ state. (a) Relationship between the monoclinic unit cell and the $2b_o$ pseudo-orthorhombic $o'$ cell defined by \autoref{eq:S_phonon_transformation}. (b,c) Calculated dispersion relations in the monoclinic and $o'$ cells, respectively.
}
\vspace{0.5cm}
\end{figure*}

To assess the dynamical stability of the $P2_1$ phase, the phonon dispersion in \autoref{fig:S_PhononP21}(b) was calculated. Although shallow imaginary frequencies occur along interpolated portions of the branches, none corresponds to a resolved soft mode at a sampled high-symmetry wave vector. In addition, the corresponding modes are acoustic rather than optical. Taken together, these features support treating the small values as numerical artifacts within the resolution of the finite-displacement calculation rather than evidence of a structural instability.

To test whether the numerical artifacts are associated with the lower symmetry and nonorthogonal lattice vectors of the monoclinic cell, the calculation was repeated in a pseudo-orthorhombic supercell. Specifically, a transformation analogous to that used in the main text (\autoref{main:eq:transformation}) was constructed:
\begin{equation}
\begin{pmatrix}
a_{o'} \\ b_{o'} \\ c_{o'}
\end{pmatrix}
=
\begin{pmatrix}
0 & 0 & 1 \\ 4 & 0 & 1 \\ 0 & 1 & 0
\end{pmatrix}
\begin{pmatrix}
a_m \\ b_m \\ c_m
\end{pmatrix}.
\label{eq:S_phonon_transformation}
\end{equation}
Under this transformation, the cell contains a $2b_o$ repeat, $b_{o'}=2b_o$, while $a_{o'}=a_o$ and $c_{o'}=c_o$. The $2b_o$ repeat is required to construct the atomic structure explicitly because otherwise the split crystallographic sites are placed unphysically close together.

\autoref{fig:S_PhononP21}(a) shows the resulting supercell. For the phonon calculation, an additional twofold repeat along $a$ was used, with the $X$ point indicated in \autoref{fig:S_PhononP21}(c). In addition, the transformed-cell calculation includes the $\Gamma$ and $A$ points of the standard monoclinic cell [\autoref{fig:S_PhononP21}(b)]. Consistent with the numerical interpretation above, no imaginary branch occurs between $\Gamma$ and $X$.

\clearpage

\section{Contributions of vibrational free energies}

\begin{figure*}[htbp]
\includegraphics[width=0.90\textwidth]{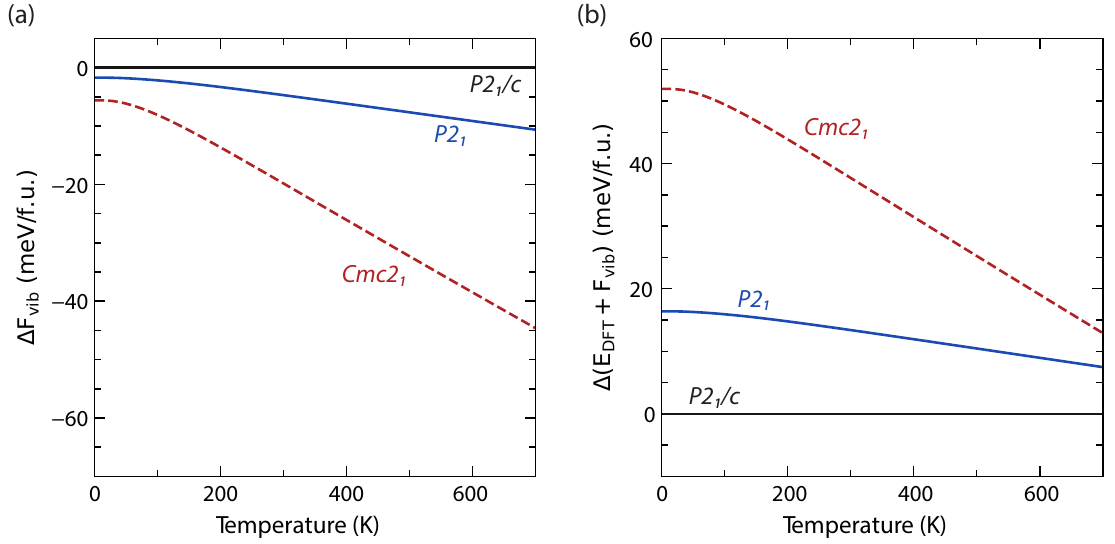}
\caption{\label{fig:S_Fvib} 
Temperature dependence of (a) the vibrational free-energy differences, $\Delta F_{\mathrm{vib}}$, and (b) the combined static DFT and vibrational free-energy differences, $\Delta(E_{\mathrm{DFT}}+F_{\mathrm{vib}})$, referenced to $P2_1/c$ at each temperature. Calculations use each phase’s relaxed 0 K structure at fixed volume and exclude non-positive phonon frequencies. The dashed $Cmc2_1$ curves represent an incomplete harmonic estimate because of its unstable phonon branch.
}
\vspace{0.5cm}
\end{figure*}

The contribution of lattice vibrations to finite-temperature phase stability is estimated by adding the harmonic vibrational free energy to the static DFT energy \cite{Shin/Lee2016,Kim/Shin2026,Huang/Rondinelli2016}, $F_i(T)=E_{\mathrm{DFT},i}+F_{\mathrm{vib},i}(T)$, where each phase $i$ is evaluated at its relaxed 0 K volume.
As shown in \autoref{fig:S_Fvib}, $F_{\mathrm{vib}}$ decreases with respect to temperature,  most strongly for $Cmc2_1$, increasingly favoring its stabilization relative to the monoclinic phases. Extending the comparison beyond the plotted range gives an apparent free-energy crossing between $Cmc2_1$ and $P2_1/c$ above 900 K, considerably higher than the experimental boundary near 520 K. However, these calculations exclude imaginary phonon frequencies and neglect anharmonicity, so this crossing does not provide a quantitative transition temperature. In particular, thermal stiffening of the initially unstable $S_2$ mode remains central to the proposed transition mechanism in the phenomenological description discussed in the main text, but is not captured by the present harmonic free-energy comparison.

\clearpage

\section{Spontaneous polarization}

To quantify the macroscopic polarization along the calculated structural pathways in LaTaO$_4$, spontaneous polarization values were computed using the Berry-phase method. In the $n=2$ Carpy-Galy structure of LaTaO$_4$, the layered lattice connectivity allows the octahedral-rotation distortion to acquire a polar character. The accompanying La displacements enhance the resulting dipole rather than representing a conventional orbital-hybridization-driven off-centering instability. Tracking the polarization quantum across the structural branches confirms the lower polarization of metastable $P2_1$ relative to fully developed $Cmc2_1$.

\begin{figure*}[htbp]
\includegraphics[width=0.99\textwidth]{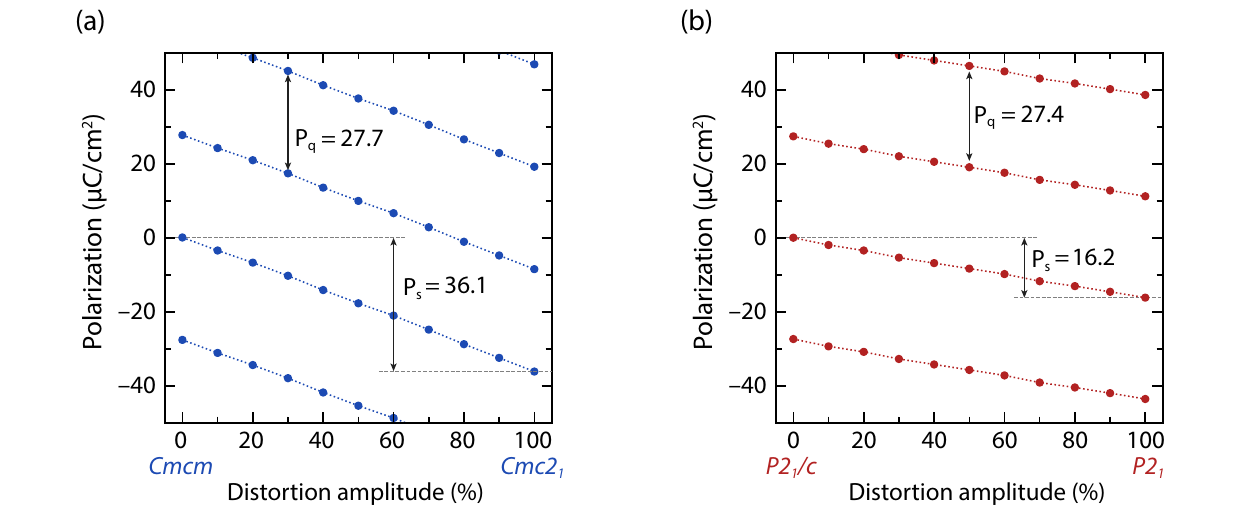}
\caption{\label{fig:P_branches}
Spontaneous polarization calculated using the Berry phase formalism. (a) Polarization branches along the displacement coordinate from nonpolar $Cmcm$ to polar $Cmc2_1$, giving a spontaneous polarization ($P_s$) of 36.1 \Ccm and a polarization quantum ($P_q$) of 27.7 \Ccm. (b) Polarization branches from the antipolar $P2_1/c$ ground state to the metastable polar $P2_1$ state. Alternating La displacements partially compensate, reducing $P_s$ to 16.2 \Ccm.
}
\vspace{0.5cm}
\end{figure*}

\clearpage

\section{Effect of spin--orbit coupling on the electronic structure}

The effect of spin--orbit coupling (SOC) on the relative phase energetics and electronic structures was investigated for \LTO. The structures were reoptimized and the electronic properties were recalculated with SOC using the PBEsol exchange--correlation functional, with all other computational settings unchanged from those described in the main text. The optimized lattice parameters, relative energies, band gaps, and spontaneous polarizations are summarized in \autoref{table:SOC_data}. Relative to the non-SOC results in \autoref{main:table:LTO_Lattice_Energy}, inclusion of SOC produces only small changes and does not alter the energetic ordering or the conclusions.

\begin{table}[htbp]
\caption{\label{table:SOC_data}
Lattice parameters (\AA), monoclinic angle $\beta$, unit-cell volume (\AA$^3$), relative energy, band gap, and polarization of the \LTO phases calculated using the PBEsol exchange--correlation functional with spin--orbit coupling. Here, $\Delta E$ is referenced to $P2_1/c$, $E_g$ is the PBEsol Kohn--Sham band gap, and $P$ is the magnitude of the spontaneous polarization. For the orthorhombic phases, $\beta=90^\circ$.
}
\centering
\begin{ruledtabular}
\begin{tabular}{r | c c c c c | r r r }
     Space group     & $a$ & $b$ & $c$ & $\beta$ &  $V$ & $\Delta E$ (meV/f.u.)  & $E_g$ (eV) & $P$ ($\mu\text{C/cm}^2$)\Bstrut \\
     \hline \Tstrut
$Cmcm$ (No. 63)      & 3.9660 & 15.352 & 5.5693 & 90.0 & 339.09 & 220.54 & 2.81 & 0 \\
$Cmc2_1$ (No. 36)    & 3.9444 & 14.673 & 5.5996 & 90.0 & 324.09 & 58.47 & 3.25 & 35.0 \\
$P2_1/c$ (No. 14)    & 7.6262 & 5.5297 & 7.7885 & 101.0 & 322.43 & 0 & 3.36 & 0 \\
$P2_1$ (No. 4)       & 7.7442 & 5.5452 & 7.7989 & 104.4 & 324.34 & 18.1 & 3.252 & 16.8 \\
\end{tabular}
\end{ruledtabular}
\end{table}
\vspace{0.0cm}

\section{Effect of exchange-correlation functionals}

\autoref{table:XC_energies} summarizes the relative phase energies and band gaps calculated using r$^2$SCAN and HSE06. Structural relaxation was performed with r$^2$SCAN, whereas HSE06 was used for single-point calculations on PBEsol-relaxed structures. Both treatments preserve the PBEsol energetic ordering $P2_1/c < P2_1 < Cmc2_1 < Cmcm$. The $P2_1$ phase lies 13.44 and 11.49 meV/f.u. above $P2_1/c$ for r$^2$SCAN and HSE06, respectively, compared with 18.10 meV/f.u. for PBEsol. Its energy separation from $Cmc2_1$ decreases from 39.44 meV/f.u. for PBEsol to 25.83 and 36.51 meV/f.u., respectively, preserving the intermediate energetic position of $P2_1$. All phases remain insulating, and the HSE06 band gaps exceed the r$^2$SCAN values by 1.00--1.06 eV.

\begin{table}[htbp]
\caption{\label{table:XC_energies}
Relative energies $\Delta E$ (meV/f.u.) and band gaps $E_g$ (eV) of the \LTO phases calculated using r$^2$SCAN and HSE06. For each functional, energies are referenced to the corresponding $P2_1/c$ phase. HSE06@PBEsol denotes HSE06 single-point calculations on PBEsol-relaxed structures.
}
\centering
\vspace{0.5em}
\setlength{\tabcolsep}{12pt}
\begin{tabular}{l|cc|cc}
\hline\hline
Space group & \multicolumn{2}{c|}{r$^2$SCAN} & \multicolumn{2}{c}{HSE06@PBEsol}\Bstrut \\
\cline{2-3}\cline{4-5}
 & $\Delta E$ (meV/f.u.) & $E_g$ (eV) & $\Delta E$ (meV/f.u.) & $E_g$ (eV)\Bstrut \\
\hline \Tstrut
$Cmcm$ (No. 63)   & 216.91 & 3.39 & 181.66 & 4.41 \\
$Cmc2_1$ (No. 36) & 39.27  & 3.83 & 48.00  & 4.88 \\
$P2_1/c$ (No. 14) & 0.00   & 3.95 & 0.00   & 4.99 \\
$P2_1$ (No. 4)    & 13.44  & 3.81 & 11.49  & 4.87 \\
\hline\hline
\end{tabular}
\end{table}
\vspace{0.0cm}

\clearpage
\section{Stability of Carpy-Galy structures relative to fergusonite and scheelite polymorphs}

The 0 K calculations identify a narrow compositional range in which the Carpy-Galy topology is energetically competitive with other $RB$O$_4$ polymorphs. Compounds with $RB$O$_4$ stoichiometry frequently crystallize as scheelite structures containing isolated tetrahedra or as fergusonite structures containing edge-connected octahedral chains, rather than forming the extended perovskite slabs characteristic of the Carpy-Galy topology. Calculated total energies across the rare-earth ($R$) and B-site ($B$) series show that the polymorph competition is sensitive to cation chemistry.

The M'-fergusonite, M-fergusonite, and scheelite structures were optimized using the PBEsol functional, PAW potentials, 550\,eV plane-wave cutoff, and convergence criteria described in the main text. Both the lattice parameters and atomic coordinates were relaxed. $\Gamma$-centered $k$-point meshes of $6\times6\times6$, $6\times3\times6$, and $6\times3\times6$ were used for M'-fergusonite, M-fergusonite, and scheelite, respectively. Initial structures followed the reported structure types cited below.

The three competing polymorphs differ in lattice symmetry and polyhedral connectivity. Scheelite is tetragonal with space group $I4_1/a$ (No.~88) and contains isolated $B$O$_4$ tetrahedra. Monoclinic symmetry breaking and the accompanying structural distortion produce the fergusonite polymorphs, in which edge-connected $B$O$_6$ octahedra form chains. M-fergusonite is the body-centered subgroup distortion of scheelite and adopts the commonly used $I2/a$ setting of $C2/c$ (No.~15), whereas M'-fergusonite is a distinct primitive monoclinic polymorph described in the $P2/a$ setting of $P2/c$ (No.~13) \cite{Wolten1967,Garg2024,Banerjee2025,Errandonea2019}. The two fergusonite polymorphs differ in the stacking of the octahedral chains along the $b$ axis: the chains are staggered in M-fergusonite but aligned in M'-fergusonite. Both chain arrangements remain structurally distinct from the extended octahedral slabs of the Carpy-Galy phase.

Among the surveyed $R$TaO$_4$ and $R$NbO$_4$ compositions ($R=\mathrm{La,Ce,Pr,Nd}$), LaTaO$_4$ is the only composition for which monoclinic Carpy-Galy $P2_1/c$ lies below the competing fergusonite and scheelite polymorphs in total energy. For every other composition examined, at least one fergusonite- or scheelite-type structure is energetically preferred over the Carpy-Galy phase. The calculations therefore identify a LaTaO$_4$-specific stability window for the $n=2$ Carpy-Galy topology among the compositions and structures examined. Here, energetic instability denotes a higher 0 K total energy relative to a competing polymorph rather than a phonon instability.

\begin{figure*}[htbp]
\centering
\includegraphics[width=0.85\textwidth]{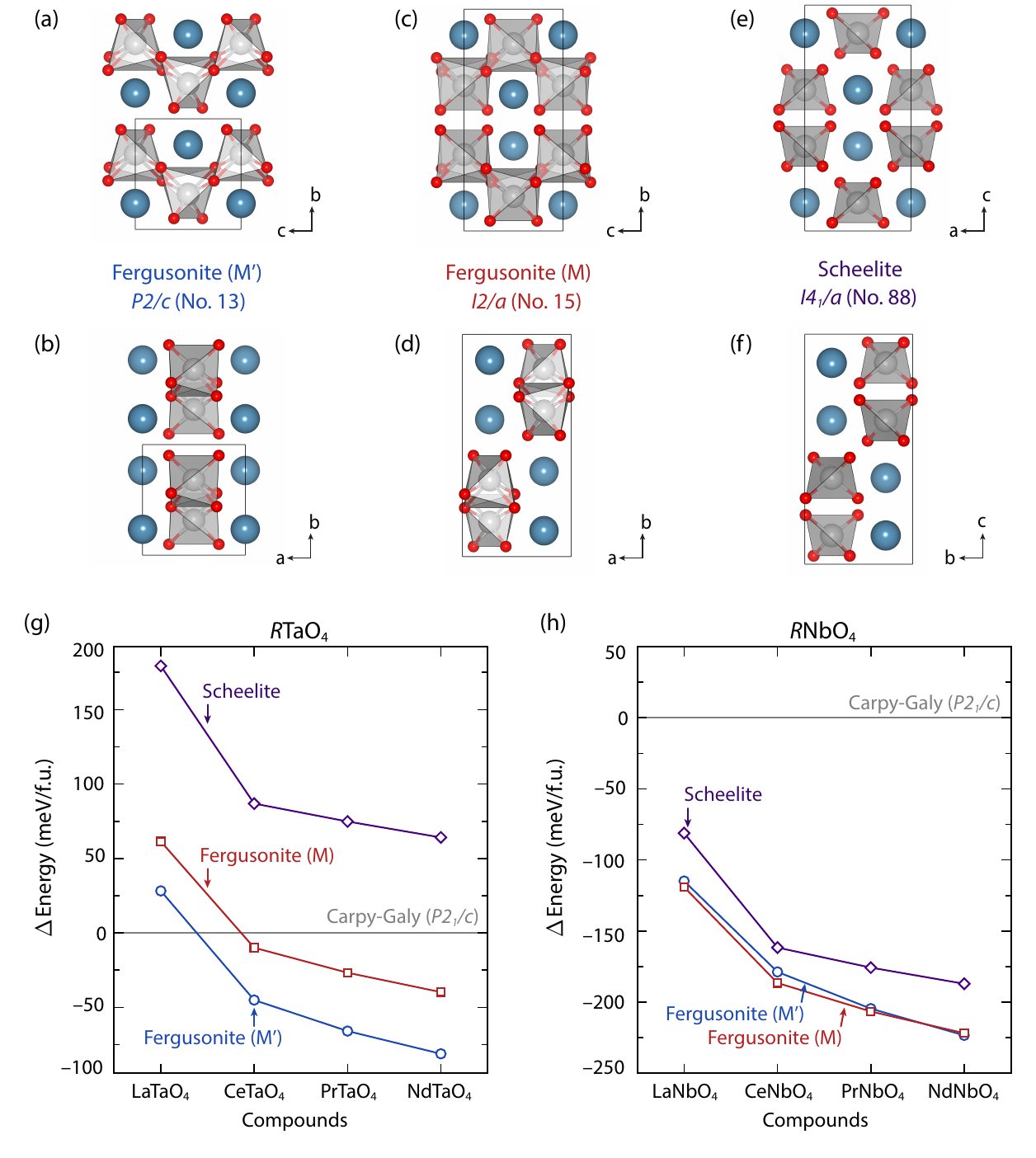}
\vspace{-2em}
\caption{\label{fig:S_Fergusonite}
Structural schematics and relative 0 K energies of competing $RB$O$_4$ polymorphs: (a,b) primitive monoclinic M'-fergusonite [$P2/a$, equivalent to $P2/c$ (No.~13)], (c,d) body-centered monoclinic M-fergusonite [$I2/a$, equivalent to $C2/c$ (No.~15)], and (e,f) tetragonal scheelite [$I4_1/a$ (No.~88)]. Blue, grey, and red spheres represent $R$, $B$ (Ta or Nb), and O atoms. (g,h) Total energy ($\Delta E$) as a function of rare-earth cation size for the $R$TaO$_4$ and $R$NbO$_4$ series, respectively. Energies are referenced to the lowest-energy Carpy-Galy polymorph, $P2_1/c$, for each composition.
}
\vspace{0.0cm}
\end{figure*}

\clearpage

\bibliographystyle{apsrev4-2}

\bibliography{StructureMap}